\documentclass[]{aastex701}

\usepackage{amsmath, amsfonts, amssymb}
\usepackage{cancel}

\usepackage[inline]{enumitem}

\usepackage{graphicx}
\usepackage{tikz}

\usepackage{makecell}

\usepackage{color}

\newcommand{\result}[1]{#1}

\begin{document}

\title{
What You Don't Know Won't Hurt You: Self-Consistent Hierarchical Inference with Unknown Follow-up Selection Strategies
}

\author[orcid=0000-0001-8196-9267]{Reed Essick}
\email[show]{essick@cita.utoronto.ca}
\affiliation{Canadian Institute for Theoretical Astrophysics, University of Toronto, 60 St. George Street, Toronto, ON M5S 3H8}
\affiliation{Department of Physics, University of Toronto, 60 St. George Street, Toronto, ON M5S 1A7}
\affiliation{David A. Dunlap Department of Astronomy, University of Toronto, 50 St. George Street, Toronto, ON M5S 3H4}

\author[orcid=0000-0002-6121-0285]{Amanda M. Farah}
\email[show]{afarah@cita.utoronto.ca}
\affiliation{Canadian Institute for Theoretical Astrophysics, University of Toronto, 60 St. George Street, Toronto, ON M5S 3H8}

\begin{abstract}
    Many astronomical surveys prompt follow-up observations, but the decision process through which candidates are selected for follow-up can be difficult to model.
    This poses a challenge when inferring properties of the intrinsic population of astrophysical sources, rather than those of the set of objects detected by the survey and often-incomplete follow-up observations.
    We alleviate this problem by demonstrating that explicitly modeling of the follow-up selection process is not required for self-consistent inference of the intrinsic population.
    Using the framework of hierarchical Bayesian inference, we show that the intrinsic population can be accurately inferred even when the decision to follow up candidates strongly correlates with latent parameters of interest.
    We provide several worked examples, showing that the precision of posterior constraints can depend on the follow-up process and that one may have to model a population of contaminants if the initial selection is imperfect.
    Our result could dramatically simplify population inference that incorporates uncoordinated follow-up from multiple observers triggered by the deluge of candidates from surveys like LSST, Gaia, and next-generation gravitational-wave interferometers.
\end{abstract}


\section{Introduction} 
\label{sec:introduction}

A fundamental goal of many astronomical surveys is to determine the properties of astrophysical populations from a catalog of detected data.
A detailed understanding of astrophysical populations can shed light on the sources' origins and occurrence rates as well as their connections to other astrophysical phenomena and fundamental physics.
However, successfully solving the inverse problem requires a detailed understanding of the survey selection, i.e., the probabilistic map that determines how likely it is for any individual astrophysical source to be included within the catalog.
Hierarchical Bayesian inference (HBI) provides a robust, principled formalism that can account for nontrivial survey selection, complicated population models, and heteroskedastic measurement uncertainties.
It posits a generative model for how catalogs are constructed, beginning with the astrophysical population and progressing through all the physical processes (including how noise is generated within detectors and how candidates are selected) to predict the observed data.
A key feature of the hierarchical framework is the conditional (in)dependencies\footnote{also called the causal structure.} assumed within the model~\citep[see, e.g., ][]{Essick:2024}.

For some survey designs, the selection function can be directly measured.
Often, this is performed through large ``injection'' campaigns in which simulated signals (injections) are drawn from a reference population and added to real data.
The combined synthetic data is processed by the same searches used to identify real candidates, and these searches assign detection statistics and statistical significances to each injection.
By comparing the detection statistics and/or significance estimates to a selection threshold, one can determine whether individual injections would end up within the catalog or not.
The entire detection process can be either deterministic or probabilistic.

Estimates of searches' selections are constructed by repeating this procedure for a large number of injections.
This has become common within gravitational-wave (GW) astronomy~\citep[see, for example, ][]{GWTC-4-RnP}, and public injection sets are curated by the LIGO-Virgo-KAGRA (LVK) Collaborations~\citep{LIGO, VIRGO, KAGRA} for each observing run~\citep{Essick:2025, O3-injections, GWTC-3-cumulative-injections, O4a-injections, GWTC-4-cumulative-injections}.
Other large collaborations similarly invest tremendous amounts of computational resources and person-power to develop injection procedures and quantify their search selection.
See, for example, work from the All-Sky Automated Survey for Supernovae~\citep[ASAS-SN, ][]{ASAS-SN-I, ASAS-SN-II, ASAS-SN-III}, the Dark Energy Spectroscopic Instrument~\citep[DESI, ][]{Bianchi:2018, Rosado-Marin:2024xte, Ross:2024nlf}, the Southern Stellar Stream Spectroscopic Survey~\citep[S5, ][]{S5} and the Dark Energy Survey~\citep[DES, ][]{S5+DES}, searches for fast radio bursts (FRBs) with the Canadian Hydrogen Intensity Mapping Experiment~\citep[CHIME, ][]{CHIME-FRB}, binary detection within Gaia catalogs~\citep{Lam:2025, ElBadry:2024}, and transiting exoplanet surveys (see, e.g.,~\citet{Burke:2015} and~\citet{Shabram:2020} among others).
Again, the common approach is to treat the selection process as algorithmic and therefore repeatable, implying that it is, at least in principle, quantifiable.

However, there are many situations in which this relatively straightforward picture does not readily apply.
In particular, it is common (and sometimes of the utmost scientific importance) to perform additional follow-up observations of candidates first identified within large surveys.
Given limited observational resources, there is often an additional decision process that determines which candidates are followed-up and which are not.
In some cases, this decision can be modeled, and the additional selection can be included within a standard hierarchical framework.
However, in many situations, the follow-up decision process can be extremely difficult to model.
For example, it could consist of humans making real-time judgments about how to allocate finite observational resources with incomplete information.
It may be extremely difficult to reproduce and, therefore, nearly impossible to measure precisely.
It is not immediately clear, then, how to incorporate follow-up observations within a self-consistent inference framework when the follow-up decision process is itself unquantifiable.
However, we expect follow-up to be informative, and, as such, we wish to include those observations when inferring the astrophysical population.

\begin{figure}
    \begin{minipage}{0.625\textwidth}
        \includegraphics[width=1.0\textwidth]{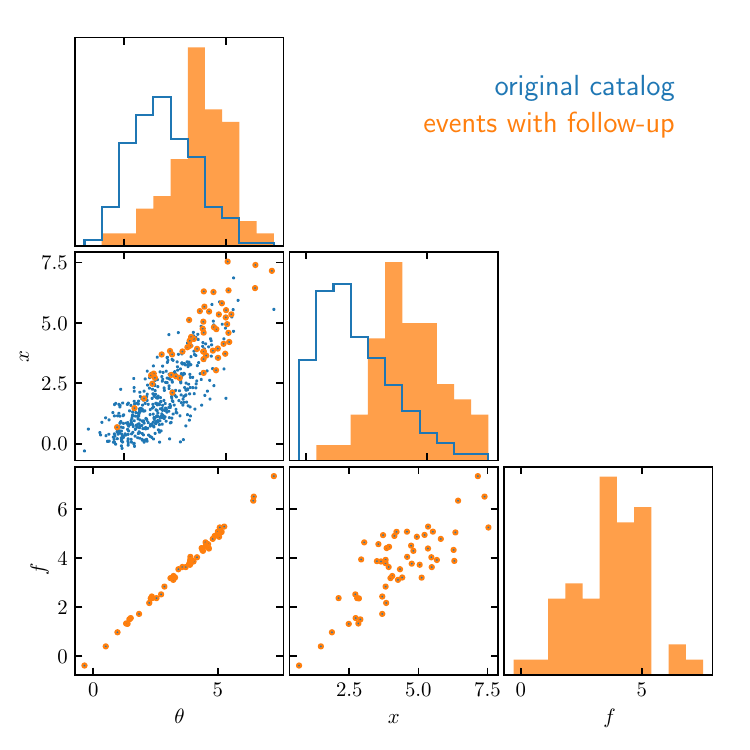}
    \end{minipage}
    \begin{minipage}{0.375\textwidth}
        \includegraphics[width=1.0\textwidth]{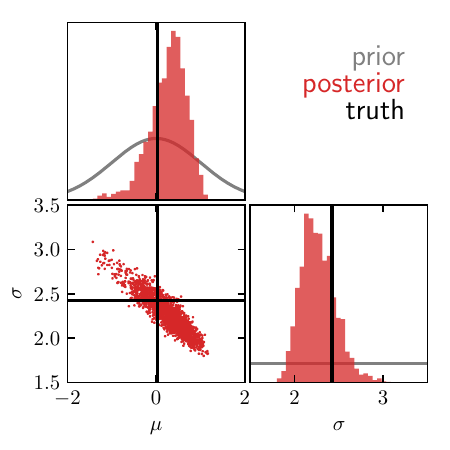}
    \end{minipage}
    \caption{
        (\emph{left}) Distributions of true event parameters and data for an example mock catalog with \result{500} detected events of which \result{56} received follow-up: (\emph{blue}) distributions of detected sources and (\emph{orange}) distributions of sources that received follow-up.
        We show distributions over the true event parameters ($\theta$), original catalog data ($x$), and follow-up data ($f$).
        See Sec.~\ref{sec:gaussian toy model} and Appendix~\ref{sec:robustness checks} for more details.
        This catalog uses a logistic follow-up probability (Eq.~\ref{eq:simplified toy model follow up}).
        Most, but not all, events with large $x$ are followed-up, and very few events with small $x$ are followed-up.
        (\emph{right}, \emph{red}) Posterior and (\emph{grey}) prior distributions over the population mean ($\mu$) and standard deviation ($\sigma$) with (\emph{black lines}) the true parameters overlaid.
    }
    \label{fig:simplified toy model}
\end{figure}

We present the (potentially surprising) result that it is not only possible but relatively straightforward to construct a self-consistent hierarchical inference that includes additional follow-up observations without explicitly modeling the follow-up decision process.
Fig.~\ref{fig:simplified toy model} illustrates this with a simulated catalog.
In this example, candidates are selected from an initial survey to create a catalog, and only a subset of events are followed-up.
The follow-up selection process does not simply follow the detected population and is correlated with the parameters of interest.
The follow-up observations are therefore a biased sample of the intrinsic distribution.
However, our framework nevertheless results in an accurate inference of the intrinsic distribution without modeling the follow-up selection process: the hyperposterior is well-centered on the true parameters.

This has immediate implications for follow-up observations of any large survey, including GW catalogs produced by the LVK, electromagnetic (EM) transient catalogs produced by the Legacy Survey of Space and Time (LSST) at the Vera C. Rubin Observatory~\citep{LSST, LSST-Cadence}, and even non-transient catalogs like those produced by Gaia~\citep{Gaia}, DESI~\citep{DESI, Bianchi:2018, Rosado-Marin:2024xte, Ross:2024nlf}, and exoplanet transit surveys (e.g., Kepler~\citep{Kepler} and TESS~\citep{TESS}).
See Table~\ref{tab:glossary} for more examples.
We show that one can obtain unbiased estimates with correct coverage for the true astrophysical population and rate of systems (rather than the observed distribution and rate) and the parameters of individual systems without modeling the process by which follow-up decisions are made.

This result is particularly timely as many large catalogs are either already producing or about to produce a higher rate of detections than could possibly be followed-up.
Understanding the impact of the inevitable decision processes that will determine which events are selected will be of utmost importance.
Our result also shows that different observers can adopt different follow-up processes without coordinating beforehand and, nevertheless, information from all of them can be trivially combined \textit{post hoc}.

We describe a general structure for hierarchical models in the presence of follow-up observations in Sec.~\ref{sec:framework}.
Fig.~\ref{fig:dag} shows the type of models we consider.
Sec.~\ref{sec:applications} provides a few astrophysical applications of this result, including a glossary that maps our notation to more concrete observables and astrophysical processes (Table~\ref{tab:glossary}).
We conclude in Sec.~\ref{sec:discussion} with discussions of the benefits and drawbacks of this approach.
In some cases, it may still be favorable to model the follow-up selection process explicitly and only consider events that were followed-up.
However, our results shows that progress can still be made when that is not possible.


\section{Statistical Framework}
\label{sec:framework}

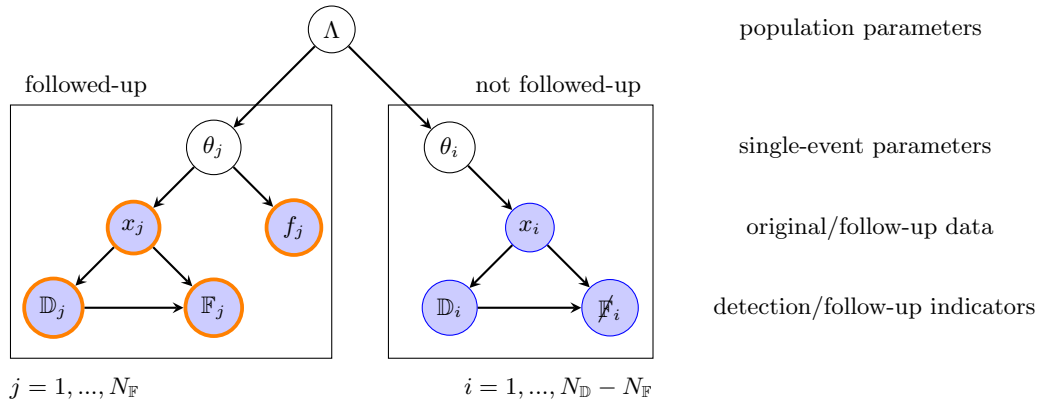
\begin{figure*}
    \begin{center}
    \begin{tikzpicture}[node distance=1.5cm]
        \tikzstyle {node} = [circle, text centered, draw=black, fill=white];
        \tikzstyle {nolnode} = [circle, text centered, draw=blue, fill=blue, fill opacity=0.2, text opacity=1.0];
        \tikzstyle {folnode} = [circle, text centered, draw=orange, fill=blue, fill opacity=0.2, text opacity=1.0, line width=0.5mm];
        \tikzstyle {arrow} = [thick, ->, >=stealth, draw=black]
        \node (L) [node] {$\Lambda$} ;
        %
        %
        \draw [draw=black] (-4.25, -1.00) rectangle (-0.00, -4.35);
        \node at (-3.40, -4.75) {$j=1,...,N_\mathbb{F}$};
        \node [rotate=0] at (-3.25, -0.75) {followed-up};
        \draw [draw=black] (+0.75, -1.00) rectangle (+4.25, -4.35);
        \node at (+3.00, -4.75) {$i=1,...,N_\mathbb{D}-N_\mathbb{F}$};
        \node [rotate=0] at (+3.00, -0.75) {not followed-up};        %
        %
        \node (tj) [node, below left of=L, xshift=-0.5cm, yshift=-0.5cm] {$\theta_j$} ;
        \node (xj) [folnode, below left of=tj] {$x_j$} ;
        \node (Dj) [folnode, below left of=xj] {$\mathbb{D}_j$} ;
        \node (Fj) [folnode, below right of=xj] {$\mathbb{F}_j$} ;
        \node (fj) [folnode, below right of=tj] {$f_j$} ;
        \node (ti) [node, below right of=L, xshift=+0.5cm, yshift=-0.5cm] {$\theta_i$} ;
        \node (xi) [nolnode, below right of=ti] {$x_i$} ;
        \node (Di) [nolnode, below left of=xi] {$\mathbb{D}_i$} ;
        \node (notFi) [nolnode, below right of=xi] {$\cancel{\mathbb{F}}_i$} ;
        %
        %
        \draw [arrow] (L) -- (tj) ;
        \draw [arrow] (tj) -- (xj) ;
        \draw [arrow] (tj) -- (fj) ;
        \draw [arrow] (xj) -- (Dj) ;
        \draw [arrow] (xj) -- (Fj) ;
        \draw [arrow] (Dj) -- (Fj) ;
        \draw [arrow] (L) -- (ti) ;
        \draw [arrow] (ti) -- (xi) ;
        \draw [arrow] (xi) -- (Di) ;
        \draw [arrow] (xi) -- (notFi) ;
        \draw [arrow] (Di) -- (notFi) ;
        %
        %
        \node (DFlabel) [right of=notFi, xshift=+2.0cm] {detection/follow-up indicators} ;
        \node (xflabel) [right of=xi, xshift=+3.0cm] {original/follow-up data} ;
        \node (tlabel) [right of=ti, xshift=+4.0cm] {single-event parameters} ;
        \node (Llabel) [right of=L, xshift=+5.5cm] {population parameters} ;
    \end{tikzpicture}
    \end{center}
    \caption{
        Directed acyclic graph (DAG) representing the type of observations we consider.
        There are $N_\mathbb{D}$ detected events in total, but only $N_\mathbb{F} \leq N_\mathbb{D}$ receive follow-up.
        Importantly, both detection ($\mathbb{D}$) and the decision to follow-up ($\mathbb{F}$) or not ($\cancel{\mathbb{F}}$) for each event depend only on the initial catalog data ($x$).
        Filled nodes correspond to observed variables and unfilled nodes denote latent variables.
        Plates (rectangular boxes) denote independently and identically distributed (i.i.d.) sets of variables.
        Colors match the convention in Fig.~\ref{fig:simplified toy model}.
        See Appendixes~\ref{sec:active learning} and~\ref{sec:related dag} for extensions and related models.
    }
    \label{fig:dag}
\end{figure*}

We consider an inhomogeneous Poisson point process with selection effects (i.e., only a subset of events are detected).\footnote{This is often called a censored or thinned process in the statistics literature, and accounting for selection effects is often called a completeness correction within the astronomy literature.}
Additionally, only a subset of detected events are followed-up.
We seek posterior probability densities for the latent single-event parameters, the shape of their distribution, and the expected number of astrophysical events conditioned on observed data, detection, and the decision to follow-up events (or not).

We model the differential rate density of the process as
\begin{equation}
    \frac{d N_E}{d \theta} = N_E p(\theta|\Lambda)
\end{equation}
where $\Lambda$ denotes parameters that describe the shape of the population and $\theta$ denotes the parameters that describe individual systems.
We assume $p(\theta|\Lambda)$ is a properly normalized probability density for all $\Lambda$, and $N_E$ is the expected number of astrophysical systems (regardless of whether they are detected): i.e., $\int d\theta\, dN_E / d\theta = N_E \int d\theta\, p(\theta|\Lambda) = N_E$.

We consider the following (hypothetically observable) data: $N$ denotes the actual number of astrophysical systems that occur, and the expected value of $N$ is $N_E$.
$N_\mathbb{D}$ denotes the number of detected systems (regardless of whether they received follow-up), and $N_\mathbb{F} \leq N_\mathbb{D}$ denotes the number of systems that received follow-up.
For each event $k$, we denote the original catalog data as $x_k$ and the follow-up data, when available, as $f_k$.
The indicators $\mathbb{D}$ ($\cancel{\mathbb{D}}$) and $\mathbb{F}$ ($\cancel{\mathbb{F}}$) denote whether an event was detected (or not) and whether an event was followed-up (or not), respectively.

Using the conditional (in)dependencies shown in Fig.~\ref{fig:dag}, we derive a joint distribution for all the latent and observed variables.
Similar to the approach in~\citet{Mandel:2018}, Appendix~\ref{sec:derivations} provides both a ``top-down'' derivation that starts with astrophysical distributions and progresses towards the detected data along with a ``bottom-up'' derivation that starts with detected data and works backwards.
Both result in the same joint distribution (Eq.~\ref{eq:main result joint}), which yields a posterior (Eq.~\ref{eq:main result post}) that does not depend on $\mathbb{F}$ (or $\cancel{\mathbb{F}}$) directly.

Our main result is a joint distribution of all the observed data (from $N_\mathbb{D}-N_\mathbb{F}$ events that were not followed-up and $N_\mathbb{F}$ events that were) and latent single-event parameters for all events conditioned on the astrophysical rate and population ($N_E, \Lambda$).
\begin{align}
    p(\{\theta, x, \mathbb{D}, \cancel{\mathbb{F}}\}_i^{N_\mathbb{D}-N_\mathbb{F}}, \{\theta, x, f, & \mathbb{D}, \mathbb{F}\}_j^{N_\mathbb{F}}, N_\mathbb{F}, N_\mathbb{D} | N_E, \Lambda) \nonumber \\
    & = N_E^{N_\mathbb{D}} e^{-N_E P(\mathbb{D}|\Lambda)} \frac{1}{(N_\mathbb{D}-N_\mathbb{F})!} \left[ \prod\limits_{i=1}^{N_\mathbb{D}-N_\mathbb{F}} P(\cancel{\mathbb{F}}|\mathbb{D}, x_{i}) P(\mathbb{D}|x_{i}) p(x_{i}|\theta_i) p(\theta_i|\Lambda) \right] \nonumber \\
    & \quad \quad \quad \quad \times \frac{1}{N_\mathbb{F}!} \left[ \prod\limits_{j=1}^{N_\mathbb{F}} P(\mathbb{F}|\mathbb{D}, x_{j}) P(\mathbb{D}|x_{j}) p(x_{j}|\theta_j) p(f_{j}|\theta_j) p(\theta_j|\Lambda) \right] \label{eq:main result joint}
\end{align}
If we additionally include a prior for $(N_E, \Lambda)$, we obtain a posterior distribution from Eq.~\ref{eq:main result joint} via
\begin{multline}
    p(\{\theta\}_i^{N_\mathbb{D}-N_\mathbb{F}}, \{\theta\}_j^{N_\mathbb{F}}, N_E, \Lambda | \{x,\mathbb{D},\cancel{\mathbb{F}}\}_i^{N_\mathbb{D}-N_\mathbb{F}}, \{x, f, \mathbb{D}, \mathbb{F}\}_j^{N_\mathbb{F}}, N_\mathbb{F}, N_\mathbb{D}) \\
        = \frac{p(\{\theta, x, \mathbb{D}, \cancel{\mathbb{F}}\}_i^{N_\mathbb{D}-N_\mathbb{F}}, \{\theta, x, f, \mathbb{D}, \mathbb{F}\}_j^{N_\mathbb{F}}, N_\mathbb{F}, N_\mathbb{D} | N_E, \Lambda) p(N_E, \Lambda)}{p(\{x, \mathbb{D}, \cancel{\mathbb{F}}\}_i^{N_\mathbb{D}-N_\mathbb{F}}, \{x, f, \mathbb{D}, \mathbb{F}\}_j^{N_\mathbb{F}}, N_\mathbb{F}, N_\mathbb{D})} ,
\end{multline}
where we condition by dividing the joint distribution by the marginal distribution: i.e., $p(A|B) = p(A, B)/p(B)$.
In our case, the marginal\footnote{also called the Bayesian evidence} is
\begin{align}
    p(\{x, & \mathbb{D}, \cancel{\mathbb{F}}\}_i^{N_\mathbb{D}-N_\mathbb{F}}, \{x, f, \mathbb{D}, \mathbb{F}\}_j^{N_\mathbb{F}}, N_\mathbb{F}, N_\mathbb{D}) \nonumber \\
        & = \left[ \frac{1}{(N_\mathbb{D}-N_\mathbb{F})!} \prod\limits_{i=1}^{N_\mathbb{D}-N_\mathbb{F}} P(\cancel{\mathbb{F}}, \mathbb{D}|x_i) P(\mathbb{D}|x_i) \right] \left[ \frac{1}{N_\mathbb{F}!} \prod\limits_{j=1}^{N_\mathbb{F}} P(\mathbb{F}|\mathbb{D}, x_j) P(\mathbb{D}|x_j) \right] \nonumber \\
        & \quad \quad \times \int dN_E d\Lambda \, p(N_E, \Lambda) N_E^{N_\mathbb{D}} e^{-N_E P(\mathbb{D}|\Lambda)} \prod\limits_{k=1}^{N_\mathbb{D}-N_\mathbb{F}} d\theta_k \, p(x_k|\theta_k)p(\theta_k|\Lambda) \prod\limits_{l=1}^{N_\mathbb{D}-N_\mathbb{F}} d\theta_l \, p(x_l|\theta_l)p(f_l|\theta_l) p(\theta_l|\Lambda)
\end{align}
Because the terms that only depend on $x$, $f$, $\mathbb{D}$, $\mathbb{F}$, and $\cancel{\mathbb{F}}$ factor out of the marginalization, we obtain a term-by-term cancellation analogous to the term-by-term cancellation in the ``physical DAG'' in~\citet{Essick:2024}.
As such, we obtain
\begin{align}
    p(\{\theta\}_i^{N_\mathbb{D}-N_\mathbb{F}}, \{\theta\}_j^{N_\mathbb{F}}, N_E, \Lambda & | \{x,\mathbb{D},\cancel{\mathbb{F}}\}_i^{N_\mathbb{D}-N_\mathbb{F}}, \{x, f, \mathbb{D}, \mathbb{F}\}_j^{N_\mathbb{F}}, N_\mathbb{F}, N_\mathbb{D}) \nonumber \\
        & \propto p(N_E, \Lambda) N_E^{N_\mathbb{D}} e^{-N_E P(\mathbb{D}|\Lambda)} \left[ \prod\limits_{i=1}^{N_\mathbb{D}-N_\mathbb{F}} p(x_{i}|\theta_i) p(\theta_i|\Lambda) \right] \left[ \prod\limits_{j=1}^{N_\mathbb{F}} p(x_{j}|\theta_j) p(f_{j}|\theta_j) p(\theta_j|\Lambda)  \right] \label{eq:main result post}
\end{align}
which does not depend on $\mathbb{F}$ (or $\cancel{\mathbb{F}}$), and we do not need to model $P(\mathbb{F}|\mathbb{D},x)$.
In other words, we do not have to explicitly model the probability that any individual event would be followed-up as long as we condition on all of the original catalog data (on which the follow-up decision is based) and all of the follow-up data.
This cancellation is related to the ``ignorability'' properties of processes that produce missing data first introduced in~\citet{Rubin:1976} and sometimes called ``missing at random'' in the statistics literature.\footnote{With some massaging and changes of variables, it is possible to consider our result as an application of the ignorability property from~\citet{Rubin:1976} applied to the distribution of detected events.}

Again, Appendix~\ref{sec:derivations} provides detailed derivations.
Appendix~\ref{sec:robustness checks} demonstrates that the posterior distribution in Eq.~\ref{eq:main result post} is correct by showing the results of extensive coverage tests with different follow-up decision processes.
Appendixes~\ref{sec:active learning} and~\ref{sec:discarding follow-up} provide related extensions.


\section{Applications}
\label{sec:applications}

\begin{table*}
    \begin{center}
    \caption{
        A glossary translating our notation into concrete variables in several observational contexts.
        This is not meant to be an exhaustive list.
        Secs.~\ref{sec:gw cosmology} and~\ref{sec:rare systems} provide worked examples.
    }
    \label{tab:glossary}
    \begin{tabular}{lllllll}
            \hline
            \hline
                & \multicolumn{1}{c}{$\Lambda$}
                & \multicolumn{1}{c}{$\theta$}
                & \multicolumn{1}{c}{$x$}
                & \multicolumn{1}{c}{$f$}
                & \multicolumn{1}{c}{$\mathbb{D}|x$}
                & \multicolumn{1}{c}{$\mathbb{F}|\mathbb{D}, x$}
                \\
            \hline
            \hline
            GW/Bright Sirens
                & cosmo. params,
                & mass, spin,
                & GW data:
                & EM data:
                & GW search
                & human-led  
                \\
            (Sec.~\ref{sec:gw cosmology})
                & volumetric rate,
                & distance, redshift,
                & est'd mass,
                & est'd
                & 
                & EM follow-up
                \\
                & shape of mass, spin,
                & location
                & spin, distance
                & redshift,
                &
                & 
                \\
                & redshift, and
                & 
                & location
                & location
                \\
                & location distrib.
                \\
            \hline
            GRB+Xray
                & volumetric rate,
                & luminosity, 
                & GBM data:
                & Xray data:
                & Fermi GBM 
                & SWIFT
                \\
            (Fermi+SWIFT)
                & luminosity, distance, 
                & distance,
                & est'd
                & est'd
                & search,  
                & slewing,
                \\
                & opening-angle,
                & opening-angle,
                & spectrum,
                & location 
                & satellite pos.
                & satellite pos.
                \\
                & \& spectra distrib.
                & spectrum,
                & location
                & 
                \\
                &
                & location
                \\
            \hline
            SN/optical 
                & type, distance,
                & type, distance,
                & photometry
                & spectra
                & image 
                & human-led
                \\
            transients
                & location, composition
                & location, 
                & (images, 
                & (abundances)
                & differencing
                & follow-up
                \\
                & distrib
                & composition
                & lightcurves)
                \\
            \hline
            Gaia astrometry
                & number of sys., 
                & period,
                & photometry
                & spectra
                & scan pattern,
                & human-led
                \\
            (Sec.~\ref{sec:rare systems})
                & shape of period, 
                & eccentricity,
                & (Gaia
                & ($\Delta$RV)
                & mag. cuts, 
                & follow-up
                \\
                & eccentricity, 
                & inclination, 
                & astrometry)
                &
                & fit cascade
                \\
                & inclination, 
                & masses, location,
                \\
                & dark obj's mass, 
                & sys. type
                \\
                & luminous obj's mass,
                \\
                & location distrib.,
                \\
                & contaminant distrib.
                \\
            \hline
            Spectroscopic
                & cosmo. params,
                & location,
                & photometry:
                & spectra:
                & mag. \& 
                & fiber collision,
                \\
            Galaxy Surveys
                & galaxy autocorr.,
                & redshift,
                & est'd color,
                & est'd redshift,
                & color cuts
                & quality cuts
                \\
            (DESI, SDSS)
                & gal. type distrib.
                & gal. type
                & location,
                & gal. type
                \\
                &
                &
                & photo-$z$
                \\
            \hline
            Exoplanet
                & orbit, host type, 
                & orbit params,
                & photometry
                & spectra, 
                & transit search
                & human-led
                \\
            transit surveys
                & planet (mass, radius, 
                & host properties,
                & (timeseries)
                & multi-band
                &
                & follow-up
                \\
                & composition) distrib.
                & planet properties
                &
                & photometry
                &
                & 
                \\
            \hline
    \end{tabular}
    \end{center}
\end{table*}

We now consider a few applications to provide more concrete examples.
Table~\ref{tab:glossary} provides a glossary relating our notation to concrete astrophysical scenarios.
We begin in Sec.~\ref{sec:gaussian toy model} with a fairly simple Gaussian toy model to demonstrate the main result.
Sec.~\ref{sec:gw cosmology} discusses how follow-up observations can provide additional information even when there are no contaminants in the original catalog.
Sec.~\ref{sec:rare systems} examines how we can characterize a (sub)population of rare systems in a catalog that contains  contaminants even when the probability of following-up a system correlates strongly with the type of system.


\subsection{Gaussian Toy Model}
\label{sec:gaussian toy model}

We now discuss the model behind Fig.~\ref{fig:simplified toy model} in more detail.
Specifically, we assume
\begin{align}
    p(\theta|\Lambda) & = \frac{1}{\sqrt{2\pi \sigma^2}} \exp\left(-\frac{(\theta - \mu)^2}{2\sigma^2}\right) \\
    p(x|\theta) & = \frac{1}{\sqrt{2\pi \sigma_x^2}} \exp\left(-\frac{(x - \theta)^2}{2\sigma_x^2}\right) \label{eq:gaussian p(x|theta)} \\
    p(f|\theta) & = \frac{1}{\sqrt{2\pi \sigma_f^2}} \exp\left(-\frac{(x - \theta)^2}{2\sigma_f^2}\right) \\
    P(\mathbb{D}|x) & = \left(1 + e^{(X_\mathbb{D}-x)/S_\mathbb{D}}\right)^{-1} \label{eq:simplified toy model P(D|x)}.
\end{align}
Here, the true parameters of the individual systems, $\theta$, are distributed according to the population model, $\Lambda = (\mu, \sigma)$.
The data observed in the catalog, $x$, are related to the true parameters through the likelihood in Eq.~\ref{eq:gaussian p(x|theta)}, where the measurement uncertainty is a Gaussian centered on the true parameter with standard deviation $\sigma_x$.
This simple model considers follow-up observations that are more precise measurements of the same parameter as the original catalog, i.e. $\sigma_f \ll \sigma_x$.
Finally, the detection probability of events in the initial survey is a logistic function of $x$ (Eq.~\ref{eq:simplified toy model P(D|x)}).
The primary advantages of this model are its simplicity and the fact that most of the integrals (e.g., $\int d\theta\, p(x,f|\theta) p(\theta|\Lambda)$) can be performed analytically, allowing for the extensive coverage tests shown in Appendix~\ref{sec:robustness checks}.

Fig.~\ref{fig:simplified toy model} shows a catalog with \result{500} detected events drawn from this model and the corresponding posterior inferred without directly modeling the follow-up probability, $P(\mathbb{F}|\mathbb{D},x)$.
When generating the mock catalog in Fig.~\ref{fig:simplified toy model}, we employ (but do not model within the inference) a logistic follow-up probability
\begin{equation} \label{eq:simplified toy model follow up}
    P(\mathbb{F}|\mathbb{D},x) = \left(1 + e^{(X_\mathbb{F}-x)/S_\mathbb{F}}\right)^{-1} \Theta(\mathbb{D})
\end{equation}
with $X_\mathbb{F}=5$ and $S_\mathbb{F}=1$.
We include a factor of $\Theta(\mathbb{D})$ to account for the fact that only events that are detected can be followed-up.\footnote{$\Theta$ is the indicator function: 1 when its argument is true and 0 otherwise.}
For this particular realization, \result{56} events received follow-up.

We see that, even though the follow-up selection algorithm does not simply randomly sample the original detected catalog and instead depends on $x$, we nevertheless infer an informative posterior that is centered on the injected value.
Appendix~\ref{sec:robustness checks} shows that this is not an accident; the posteriors always have proper coverage (i.e., they are correctly calibrated) independent of the follow-up strategy.
Appendix~\ref{sec:robustness checks} considers many different follow-up strategies, including several examples with both variable and fixed numbers of follow-up observations.


\subsection{Bright Sirens (multi-messenger follow-up without contaminants)}
\label{sec:gw cosmology}

\begin{figure}
    \includegraphics[width=0.5\textwidth]{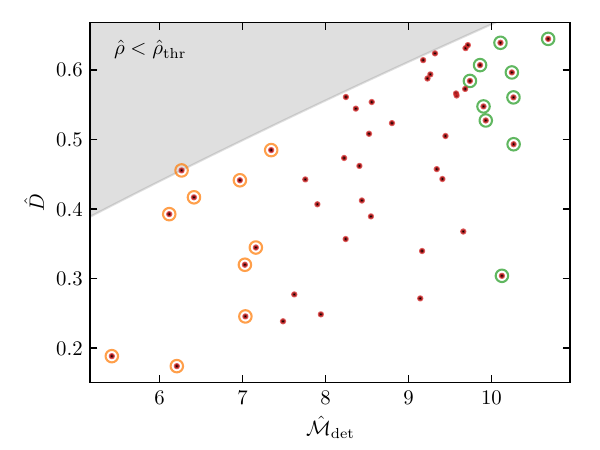}
    \includegraphics[width=0.5\textwidth]{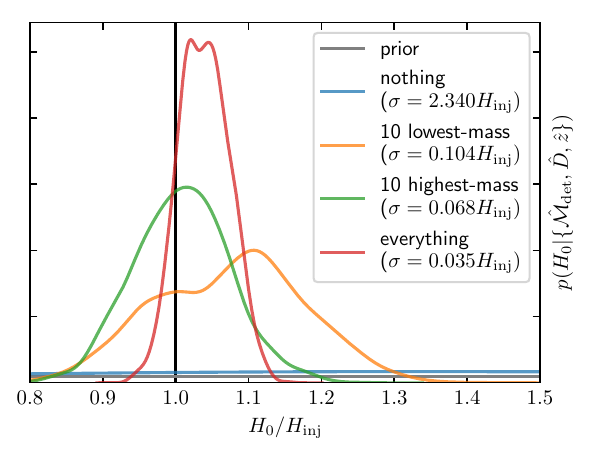}
    \caption{
        (\emph{left}) Estimated detector-frame masses ($\hat{\mathcal{M}}_\mathrm{det}$) and distances ($\hat{D}$) for individual events within a mock catalog of \result{50} detected events.
        Events that are followed-up by different schemes are circled in the corresponding color.
        The grey shaded region denotes non-detections ($\hat{\rho} \leq \hat{\rho}_\mathrm{thr}$).
        (\emph{right}) Posteriors for the Hubble parameter ($H_0$, scaled by the injected value $H_\mathrm{inj}$) using the same catalog but assuming different follow-up strategies: (\emph{blue}) nothing is followed-up, (\emph{red}) everything is followed-up, (\emph{orange}) the \result{10} events with the smallest $\hat{\mathcal{M}}_\mathrm{det}$ are followed-up, and (\emph{green}) the \result{10} events with the largest $\hat{\mathcal{M}}_\mathrm{det}$ are followed-up.
        Posterior standard deviations for each case are listed in the legend.
        We also show (\emph{gray}) the prior, which is uniform and extends far beyond any of the posteriors.
        Note that the prior and the posterior when we follow-up nothing are both much broader than the posteriors that include follow-up, although the posterior without follow-up is still informative compared to the prior (i.e., the spectral sirens measurement still carries information).
    }
    \label{fig:bright sirens}
\end{figure}

Bright sirens, or constraints on the Hubble expansion derived from joint GW and EM observations of the same compact binary coalescence, are often used as an example of the promise of multi-messenger astronomy.
In the ideal scenario,\footnote{which has only occurred once to-date~\citep{GW170817, GW170817-MMA}.} a compact binary merger observed in GWs (with data $x$) is followed-up with EM observations (producing data $f$).
The GW data, using waveforms predicted by general relativity, directly constrain the distance to the source but do not necessarily contain information about the redshift.\footnote{Redshift information can be accessed from the GW data through knowledge about the source-frame distribution of the mass, spin, and tides~\citep[e.g., ][]{Taylor:2011, Mali:2024, Tong:2025xvd, Messenger:2011gi}, sometimes called spectral, spinning, and love sirens, respectively.}
EM follow-up can identify a host galaxy, and a precise redshift can be estimated from that galaxy's spectrum.
The ratio of the estimated redshift and the luminosity distance provides an estimate for the Hubble parameter ($H_0$).
Information about $H_0$ from multiple events is combined through hierarchical inference.
See, for example, \citet{Schutz:1986}, \citet{Gair:2023}, and~\citet{Gray:2023}.

Even though the GW selection effects are well-understood and precisely measured~\citep[see, e.g.,][]{Essick:2023, Essick:2025, Essick:2026}, the procedure by which follow-up observations are obtained is not.
Indeed, the decision to conduct follow-up observations is often made in real time by observers with incomplete information, usually in the middle of the night~\citep[at least in the Western hemisphere; ][]{Essick:2026}.
That process is almost certainly not deterministic or easily predictable, and, because humans are involved, it is impractical to directly characterize it through repeated simulations.
Therefore, modeling $P(\mathbb{F}|\mathbb{D}, x)$ can be, for all intents and purposes, impossible.

Furthermore, the GW selection can be made strict enough that the purity of the initial sample is relatively high (i.e., we can be confident that everything in the original catalog is a real GW from a compact binary).

Bright sirens therefore constitute an ideal situation in which we can demonstrate unbiased inference even in the presence of unknown follow-up decision processes.
With the associations in Table~\ref{tab:glossary}, we use Eq.~\ref{eq:main result post} to draw samples from the hyperposterior over $H_0$ with a similar model to~\citet{Farah:2024}.
Fig.~\ref{fig:bright sirens} shows an example, and Appendix~\ref{sec:bright sirens model} describes our model in more detail.

Fig.~\ref{fig:bright sirens} shows estimates of the detector-frame mass ($\hat{\mathcal{M}}_\mathrm{det}$) and distance ($\hat{D}$) which are based on the initial catalog data ($x$) for a catalog of \result{$N_\mathbb{D}=50$} detected events.
We consider several follow-up strategies: either nothing is followed-up~\citep[and the $H_0$ is purely a spectral sirens measurement; e.g., ][]{Taylor:2011, Mali:2024, Farah:2024}, the \result{10} events with the smallest $\hat{\mathcal{M}}_\mathrm{det}$ are followed-up, the \result{10} events with the largest $\hat{\mathcal{M}}_\mathrm{det}$ are followed-up, or all events are followed-up.
In each case, a different set of events is chosen for follow-up, but we never explicitly model $P(\mathbb{F}|\mathbb{D}, x)$ within the posterior.
Constraints that include follow-up are significantly tighter than those that just use spectral sirens measurement, as expected~\citep{Chen:2017}, but we always find posteriors that are consistent with the injected value.

While we always recover informative posteriors relative to the prior, the precision of the constraint can depend on the follow-up strategy.
In our toy model, changing the follow-up strategy can reduce the size of the posterior standard deviation by as much as a factor of \result{$\sim 1.5$} while following up the same number of events, and following up all the events (increasing $N_\mathbb{F}$ from \result{10 to 50}) only reduces the posterior standard deviation by an additional factor of \result{$\sim 1.9$}.
Clearly, some events are more informative than others.
That is, even though we do not have to explicitly model $P(\mathbb{F}|\mathbb{D}, x)$ within the posterior, analysts can still improve the constraining power of finite observational resources by carefully selecting follow-up strategies.
For an example of optimizing follow-up strategies, see~\citet{LSST-Cadence}.

It is worth noting that expressions like parts of Eq.~\ref{eq:main result post} are commonly used in the GW cosmology community~\citep{Gray:2023}, often with the simple justification that the noise in the GW and EM observations is independent and therefore the single-event likelihoods multiply.
To the best of our knowledge, though, there has not previously been a comprehensive discussion of whether neglecting the follow-up decision process within the inference would cause biases, although Appendixes~\ref{sec:discarding follow-up} and~\ref{sec:related dag} discusses connections between our work and approaches that condition on both GW and EM detection.
Our work clarifies when one can avoid modeling $P(\mathbb{F}|\mathbb{D}, x)$.


\subsection{Searches for Rare Systems (follow-up with contaminants)}
\label{sec:rare systems}

\begin{figure*}
    \includegraphics[width=1.0\textwidth]{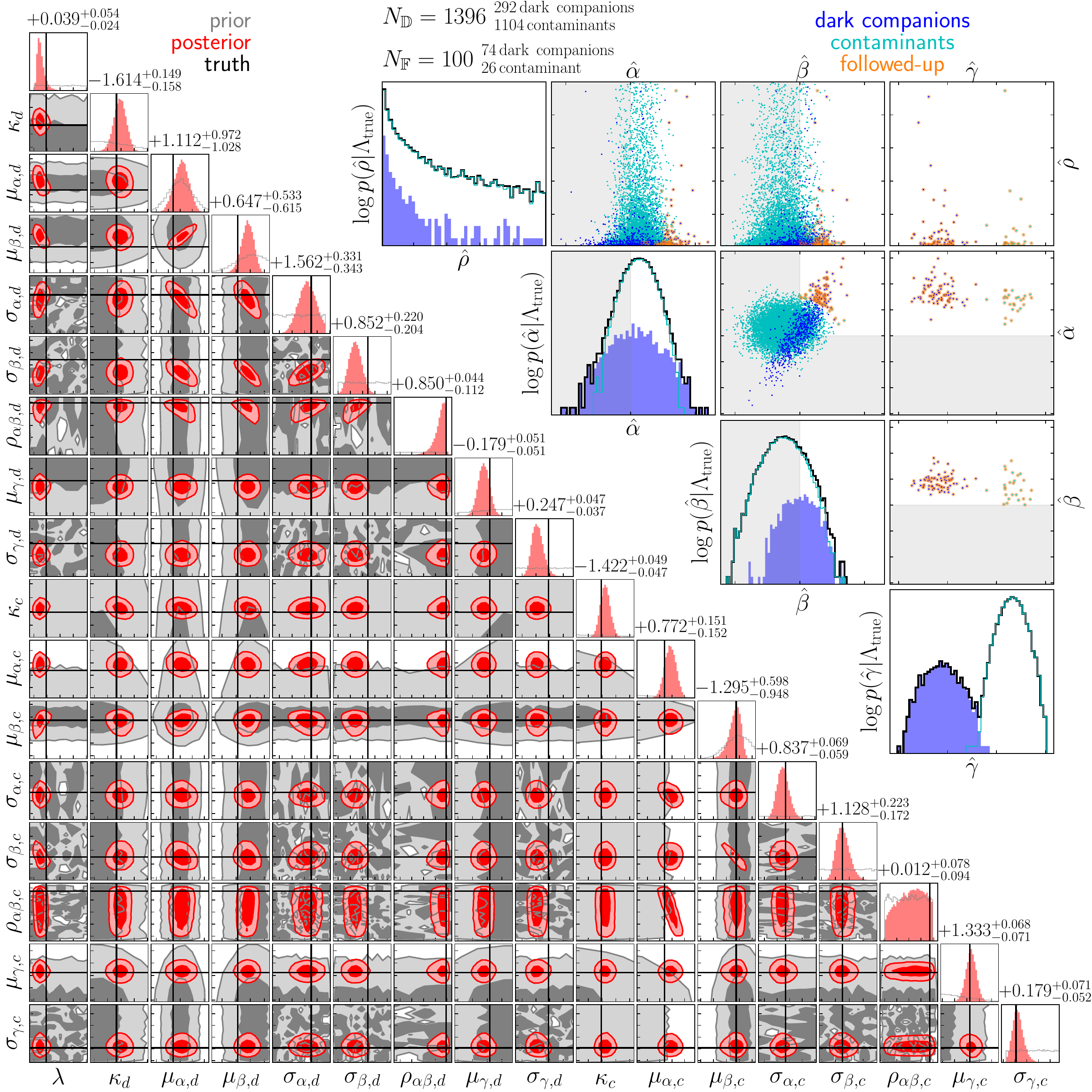}
    \caption{
        (\emph{bottom left}) Distributions of population parameters (see Appendix~\ref{sec:gaia model}) for an example catalog with \result{1396} candidates of which only the \result{100} with the largest $\hat{\alpha}^2 + \hat{\beta}^2$ were followed-up.
        Like Fig.~\ref{fig:simplified toy model}, we show the (\emph{grey}) prior, (\emph{red}) posterior, and (\emph{black}) true parameters.
        Shading denotes the \result{50\%} and \result{90\%} credible regions in the joint distributions.
        Titles of each column list the median and 90\% symmetric posterior credible region for the corresponding marginal distribution.
        Importantly, even though the fraction of systems with dark companions is only \result{$21\%$} of the original catalog and $\result{74\%}$ of the systems selected for follow-up, we nevertheless infer an astrophysical mixing fraction of \result{$\lambda = 3.9^{+5.4}_{-2.4}\%$} that is consistent with the truth (\result{$\lambda_\mathrm{true} = 7.4\%$}).
        (\emph{top right}) Distributions of observed data ($\hat{\rho}, \hat{\alpha}, \hat{\beta}$ from the original catalog and $\hat{\gamma}$ from follow-up).
        Marginal distributions show the (\emph{dark blue}) distribution of systems with dark companions, (\emph{light blue}) distribution of contaminants, and (\emph{black}) total distribution.
        Grey shading denotes regions that were not selected as interesting (i.e., detected ($\mathbb{D}$) candidates only lie in unshaded regions) and orange circles highlight the events that were followed-up.
    }
    \label{fig:rare systems}
\end{figure*}

Many catalogs contain contaminants (either other types of astrophysical systems or \textit{bona fide} noise) that may not be easily excluded by an initial selection based on the original catalog.
Follow-up is then performed to identify the true nature of each system and/or to better characterize its properties.
As shown in Table~\ref{tab:glossary}, this commonly occurs in photometric surveys that prompt spectroscopic follow-up~\citep[e.g., ][]{ASAS-SN-I, DESI, Gaia}.

We are interested in combining all initial catalog data and all follow-up observations to learn about the entire population and the individual systems simultaneously.
This is naturally accomplished by modeling the overall population as a mixture of subpopulations
\begin{equation}
    p(\theta|\Lambda) = \sum\limits_a \lambda_a p(\theta|\Lambda_a)
\end{equation}
with the constraint $\sum_a \lambda_a = 1$ and the understanding that $\Lambda$ is the concatenation of all subpopulation parameters (including the mixing fractions): $\Lambda = \bigoplus_a \{\lambda_a, \Lambda_a\}$.
Often, we may be particularly interested in the mixture weights since they determine the relative proportions of the overall population that belong to each class: i.e., $\lambda_a = p(a|\Lambda)$.

Based on searches for dark companions to main sequence stars observed by Gaia~\citep[e.g., ][]{ElBadry:2023a, ElBadry:2023b, Chakrabarti:2023, Muller-Horn:2025}, we consider a situation in which  follow-up selection based on the initial data correlates strongly with a single subpopulation (i.e., follow-up is efficiently finds interesting systems).
Specifically, we assume there are two subpopulations: systems with dark companions ($d$) and contaminants ($c$).
Events are detected and selected as potentially interesting using the initial catalog data, which provides noisy estimates of three parameters: $\hat{\rho}$, $\hat{\alpha}$, and $\hat{\beta}$.
These could correspond to the systems luminosity, \texttt{ruwe}~\citep{Lindegren:2018}, and \texttt{amrf}~\citep{Shahaf:2019}, respectively, or whatever estimates from the initial catalog are used identify interesting systems.
For the systems that are followed-up, we also observe $\hat{\gamma}$, which is a strong (but not perfect) indicator of the system's type (i.e., the $p(\gamma|a)$ do not overlap much for different subpopulations).

Fig.~\ref{fig:rare systems} demonstrates our result for a single catalog realization.
See Appendix~\ref{sec:gaia model} for more details and coverage tests.
Indeed, we do not only learn about the population of interest (dark companions) but also about the contaminant population.
In principle, this could be used to make better decisions about which systems to follow-up in the future, and Appendix~\ref{sec:active learning} discusses how our result can be applied in that context, too.

Perhaps most reassuringly, even though the fraction of systems with dark companions in the initial detected catalog is \result{21\%} and the fraction in the subset that is followed-up is \result{74\%}, both of which are different from the true astrophysical fraction, \result{$\lambda_\mathrm{true} = 7.4\%$}, we correctly infer the astrophysical fraction as \result{$\lambda = 3.9^{+5.4}_{-2.4}\%$}.
This confirms that we can learn about the size of rare subpopulations in the presence of significant contamination and only probabilistic classification based on incomplete and selective follow-up observations.

However, a key part of this approach is to model both the subpopulation of interesting systems and the subpopulation of contaminants.
Presumably, the distribution of contaminants may be quite complicated and difficult to model from first principles.
It is unclear how misspecifying either subpopulation model would affect the inferred population or whether a flexible contaminant population could inadvertently ``eat'' the interesting subpopulation.\footnote{This does not appear to happen in practice because of the significant Occam penalty incurred by flexible models. See, for example,~\citet{Godfrey:2023}.}
The exact extent of these issues may be problem-specific, and we leave detailed studies to future work.


\section{Discussion}
\label{sec:discussion}

We have shown the surprising result that it is possible to construct a self-consistent hierarchical inference without explicitly modeling the decision process which governs follow-up observations.
This very general framework alleviates a major challenge in obtaining unbiased occurrence rates and population properties of rare astrophysical phenomena, and it can be readily applied to many situations (see Table~\ref{tab:glossary} for examples).
As mentioned in Sec.~\ref{sec:framework}, this is related to the ignorability properties of processes that produce missing data first introduced in~\citet{Rubin:1976} and sometimes called ``missing at random'' in the statistics literature.
However, our result is somewhat more subtle, as we still have to model the underlying detection process $P(\mathbb{D}|\Lambda)$ in Eq.~\ref{eq:main result post} even though detection naively appears to similarly satisfy the missing at random condition (i.e., only depends on the observed data and not latent parameters of individual systems).

We provide additional derivations and related results in the appendices.
Appendix~\ref{sec:derivations} provides both ``top-down'' and ``bottom-up'' derivations of our main results (Eqs.~\ref{eq:main result joint} and~\ref{eq:main result post}).
Appendix~\ref{sec:robustness checks} numerically demonstrates proper coverage for a variety of follow-up strategies using the Gaussian model from Sec.~\ref{sec:gaussian toy model}.
Appendixes~\ref{sec:active learning},~\ref{sec:discarding follow-up}, and~\ref{sec:related dag} discuss more complicated conditional dependencies between $\mathbb{F}$ and observed data, how discarding follow-up data based on the same data spoils the term-by-term cancellation, and the relationship between our main result and a related DAGs considered in the literature, respectively.
Appendixes~\ref{sec:bright sirens model} and~\ref{sec:gaia model} describe the models used in Secs.~\ref{sec:gw cosmology} and~\ref{sec:rare systems}.

We consider a few toy examples in detail in Sec.~\ref{sec:applications}, but it is worth spelling out the implications for large surveys like LSST~\citep{LSST, LSST-Cadence} and/or future Gaia~\citep{Gaia} data releases again.

Considering LSST, one still has has carefully characterize the behavior of individual alert brokers (i.e., $P(\mathbb{D}|\Lambda)$) but does not have to characterize how observers select alerts from those brokers for additional follow-up.
This also obviates the need for observers to coordinate \textit{a priori} (beyond optimizing follow-up resources).
That is, each observer can independently select alerts for follow-up, and follow them up to differing degrees, without needing to model any of those decision processes as long as the inference is based on the entire catalog produced by the broker.

Similarly for Gaia, one still must model the cascade of processes that produced astrometric fits of varying complexity from the Gaia scanning pattern in Gaia DR3~\citep{Gaia-DR3} along with any additional selections based on those fits to capture $P(\mathbb{D}|\Lambda)$.
See~\citet{Lam:2025} and~\citet{ElBadry:2024} for examples.
However, once a list of candidates is established, follow-up can proceed in any manner.

In both cases, observers can opportunistically follow-up candidates whenever telescopes are available without worrying about the exact scheduling or completeness of follow-up.

However, when possible, it may still be preferable to model follow-up decision processes directly and only consider candidates based on follow-up observations.
Although using all the available data should provide the most constraining power, there may be other practical limitations that analysts should consider.
Specifically, if initial detection cuts produce catalogs with significant contamination, it is unclear how much model misspecification of the (possibly poorly understood) contaminant distribution could affect the inference of the subpopulation of interest.
If one adopts an extremely flexible model for the contaminant distribution, this may ``eat'' part of the distribution of interesting systems as well.
As such, modeling both simultaneously may only be advantageous if this significantly simplifies the selection and one can confidently assert some differences between the contaminant and target distributions \textit{a priori}.
The latter usually boils down to assumptions about the ``smoothness'' of the distributions (i.e., contaminants are broadly distributed and targets are narrowly distributed).

Additionally, even though one does not have to explicitly model $P(\mathbb{F}|\mathbb{D},x)$ within the posterior, which follow-up strategies optimize the expected precision at fixed observational effort remains an open question.
Thus, analysts wishing to implement algorithmic follow-up may still wish to explicitly model and test the effects of different follow-up strategies.
Again, see~\citet{LSST-Cadence} for an example.
Nonetheless, if algorithmic follow-up is impractical (or simply not achieved in practice), our result shows that self-consistent hierarchical inference is not only still possible but relatively straightforward.


\begin{acknowledgments}
    We are sincerely grateful for conversations with Biprateep Dey, Maria Drout, Maya Fishbach, Neige Frankel, Tobias G\'{e}ron, Ren\'{e}e Hlo\v{z}ek, Daniel E. Holz, Alex Laroche, Johanna M\"{u}ller-Horn, Hans-Walter Rix, Josh Speagle, Connor Stone, and Aditya Vijaykumar during the preparation of this manuscript.
    
    This research was supported by the Natural Sciences and Engineering Research Council of Canada (NSERC).
    R.E. is supported by NSERC Grant RGPIN-2023-03346, and A.M.F. is supported by NSERC Grant DIS-2022-568580.

    This research was also supported in part by grant NSF PHY-2309135 to the Kavli Institute for Theoretical Physics (KITP) and made use of the Astrophysics Data System, which is funded by NASA under Cooperative Agreement 80NSSC21M00561.
\end{acknowledgments}

\software{
\texttt{jax}~\citep{Jax:2018}, \texttt{matplotlib}~\citep{Matplotlib:2007}, \texttt{numpy}~\citep{Numpy:2020}, \texttt{numpyro}~\citep{Numpyro:2019}, and \texttt{scipy}~\citep{Scipy:2020}.
}


\appendix


\section{Derivations of the main result}
\label{sec:derivations}

Appendix~\ref{sec:top-down} provides a ``top-down'' derivation of the main result (Eqs.~\ref{eq:main result joint} and~\ref{eq:main result post}), and Appendix~\ref{sec:bottom-up} provides a complementary ``bottom-up'' derivation.


\subsection{Top-Down derivation}
\label{sec:top-down}

Begin by considering a Poisson distribution for the number of astrophysical systems ($N$) given an expected number ($N_E$)
\begin{equation}
    P(N|N_E) = \frac{1}{N!} N_E^N e^{-N_E} . \label{eq:poisson N|NE}
\end{equation}
Next, consider the binomial probability of detecting $N_\mathbb{D}$ events out of $N$ with probability $P(\mathbb{D}|\Lambda) = \int dx d\theta P(\mathbb{D}|x) p(x|\theta) p(\theta|\Lambda)$ and $P(\cancel{\mathbb{D}}|\Lambda) = 1 - P(\mathbb{D}|\Lambda)$.
\begin{equation}
    P(N_\mathbb{D}|N, P(\mathbb{D}|\Lambda)) = \left( \begin{matrix} N \\ N_\mathbb{D} \end{matrix}\right) P(\mathbb{D}|\Lambda)^{N_\mathbb{D}} P(\cancel{\mathbb{D}}|\Lambda)^{N-N_\mathbb{D}} . \label{eq:binomial ND|N}
\end{equation}
Similarly, write the probability of following-up $N_\mathbb{F}$ out of $N_\mathbb{D}$ detected systems with probability $P(\mathbb{F}|\mathbb{D},\Lambda) = P(\mathbb{F},\mathbb{D}|\Lambda)/P(\mathbb{D}|\Lambda) = \int dx d\theta P(\mathbb{F}|\mathbb{D},x) P(\mathbb{D}|x) p(x|\theta) p(\theta|\Lambda) / P(\mathbb{D}|\Lambda)$ as
\begin{equation}
    P(N_\mathbb{F}|N_\mathbb{D}, P(\mathbb{F}|\mathbb{D},\Lambda)) = \left(\begin{matrix} N_\mathbb{D} \\ N_\mathbb{F} \end{matrix}\right) P(\mathbb{F}|\mathbb{D},\Lambda)^{N_\mathbb{F}} P(\cancel{\mathbb{F}}|\mathbb{D},\Lambda)^{N_\mathbb{D}-N_\mathbb{F}} . \label{eq:binomial NF|ND}
\end{equation}
We can construct a joint distribution over $N$, $N_\mathbb{D}$, and $N_\mathbb{F}$ by multiplying Eqs.~\ref{eq:poisson N|NE},~\ref{eq:binomial ND|N}, and~\ref{eq:binomial NF|ND}.
This yields
\begin{align}
    P(N, N_\mathbb{D}, N_\mathbb{F}|N_E,\Lambda)
        & = P(N_\mathbb{F}|N_\mathbb{D}, P(\mathbb{F}|\mathbb{D},\Lambda)) P(N_\mathbb{D}|N, P(\mathbb{D}|\Lambda)) P(N|N_E) \nonumber \\
        & = \frac{N_E^N e^{-N_E}}{(N-N_\mathbb{D})! (N_\mathbb{D}-N_\mathbb{F})! N_\mathbb{F}!} P(\cancel{\mathbb{D}}|\Lambda)^{N-N_\mathbb{D}} \left[ P(\cancel{\mathbb{F}}|\mathbb{D},\Lambda) P(\mathbb{D}|\Lambda) \right]^{N_\mathbb{D}-N_\mathbb{F}} \left[ P(\mathbb{F}|\mathbb{D},\Lambda) P(\mathbb{D}|\Lambda) \right]^{N_\mathbb{F}} .
\end{align}
Now, consider the joint distribution of the true parameters ($\theta$) and the observed data ($x, f$) for $N_\mathbb{F}$ events that are both detected and followed-up (i.e. conditioned on $N_\mathbb{F}$, $\mathbb{D}$, and $\mathbb{F}$).
Construct this from the joint distribution (Fig.~\ref{fig:dag}) as follows.
\begin{align}
    p(\{\theta, x, f\}_j^{N_\mathbb{F}} | \{\mathbb{D}, \mathbb{F}\}_j^{N_\mathbb{F}}, N_\mathbb{F}, \Lambda)
        & = \frac{p(\{\theta, x, f, \mathbb{D}, \mathbb{F}\}_j^{N_\mathbb{F}} | N_\mathbb{F}, \Lambda)}{p(\{\mathbb{D}, \mathbb{F}\}_j^{N_\mathbb{F}}, | N_\mathbb{F}, \Lambda)} \nonumber \\
        & = \left(\frac{1}{P(\mathbb{F}|\mathbb{D},\Lambda) P(\mathbb{D}|\Lambda)}\right)^{N_\mathbb{F}} \prod\limits_{j=1}^{N_\mathbb{F}} P(\mathbb{F}|\mathbb{D}, x_{j}) P(\mathbb{D}|x_{j}) p(x_{j}|\theta_j) p(f_{j}|\theta_j) p(\theta_j|\Lambda) , \label{eq:single-event|N}
\end{align}
because $P(\{\mathbb{D},\mathbb{F}\}_j^{N_\mathbb{F}}|N_\mathbb{F},\Lambda) = [P(\mathbb{F}|\mathbb{D},\Lambda) P(\mathbb{D}|\Lambda)]^{N_\mathbb{F}}$.
There are similar expressions for the sets of $N_\mathbb{D}-N_\mathbb{F}$ events that are detected but not followed-up and the $N-N_\mathbb{D}$ events that are not detected.
Putting the distributions for all our variables, the number of events and the distribution over ($\theta, x, f$) for each set, together yields
\begin{align}
    p(\{\theta, x, f, \cancel{\mathbb{D}}, \cancel{\mathbb{F}}\}_k^{N-N_\mathbb{D}}, \{\theta, x, f, & \mathbb{D}, \cancel{\mathbb{F}}\}_i^{N_\mathbb{D}-N_\mathbb{F}}, \{\theta, x, f, \mathbb{D}, \mathbb{F}\}_j^{N_\mathbb{F}}, N_\mathbb{F}, N_\mathbb{D}, N | N_E, \Lambda) \nonumber \\
    & = N_E^N e^{-N_E} \frac{1}{(N-N_\mathbb{D})!} \left[ \prod\limits_{k=1}^{N-N_\mathbb{D}} P(\cancel{\mathbb{D}}|x_{k}) p(x_{k}|\theta_k) p(f_{k}|\theta_k) p(\theta_k|\Lambda) \right] \nonumber \\
\
    & \quad \quad \quad \quad \times \frac{1}{(N_\mathbb{D}-N_\mathbb{F})!} \left[ \prod\limits_{i=1}^{N_\mathbb{D}-N_\mathbb{F}} P(\cancel{\mathbb{F}}|\mathbb{D}, x_{i}) P(\mathbb{D}|x_{i}) p(x_{i}|\theta_i) p(f_{i}|\theta_i) p(\theta_i|\Lambda) \right] \nonumber \\
    & \quad \quad \quad \quad \times \frac{1}{N_\mathbb{F}!} \left[ \prod\limits_{j=1}^{N_\mathbb{F}} P(\mathbb{F}|\mathbb{D}, x_{j}) P(\mathbb{D}|x_{j}) p(x_{j}|\theta) p(f_{j}|\theta_j) p(\theta_j|\Lambda) \right]
\end{align}
where $P(\cancel{\mathbb{F}}, \cancel{\mathbb{D}}|\Lambda) = P(\cancel{\mathbb{D}}|\Lambda)$ because $P(\cancel{\mathbb{F}}|\cancel{\mathbb{D}},\Lambda) = 1$ (only detected events are followed-up) and all the factors of $P(\mathbb{F},\mathbb{D}|\Lambda)$, $P(\mathbb{D},\cancel{\mathbb{F}}|\Lambda)$, and $P(\mathbb{D}|\Lambda)$ have cancelled.
This is our starting point: a joint distribution for every possible outcome allowed by our model.

Now, marginalize over the variables we do not have access to. These are ($\theta_k, x_{k}, f_{k}$) for the $N-N_\mathbb{D}$ events that are not detected and $f_{i}$ for the $N_\mathbb{D}-N_\mathbb{F}$ events that are detected but not followed-up.
\begin{align}
    p(\{\cancel{\mathbb{D}}, \cancel{\mathbb{F}}\}_k^{N-N_\mathbb{D}}, \{\theta, x, \mathbb{D}, \cancel{\mathbb{F}}\}_i^{N_\mathbb{D}-N_\mathbb{F}}, & \{\theta, x, f, \mathbb{D}, \mathbb{F}\}_j^{N_\mathbb{F}}, N_\mathbb{F}, N_\mathbb{D}, N | N_E, \Lambda) \nonumber \\
    & = N_E^{N} e^{-N_E} \frac{1}{(N-N_\mathbb{D})!} \left[ P(\cancel{\mathbb{D}}|\Lambda) \right]^{N-N_\mathbb{D}} \nonumber \\
    & \quad \quad \quad \quad \times \frac{1}{(N_\mathbb{D}-N_\mathbb{F})!} \left[ \prod\limits_{i=1}^{N_\mathbb{D}-N_\mathbb{F}} P(\cancel{\mathbb{F}}|\mathbb{D}, x_{i}) P(\mathbb{D}|x_{i}) p(x_{i}|\theta_i) p(\theta_i|\Lambda) \right] \nonumber \\
    & \quad \quad  \quad \quad \times \frac{1}{N_\mathbb{F}!} \left[ \prod\limits_{j=1}^{N_\mathbb{F}} P(\mathbb{F}|\mathbb{D}, x_{j}) P(\mathbb{D}|x_{j}) p(x_{j}|\theta_j) p(f_{j}|\theta_j) p(\theta_j|\Lambda) \right]
\end{align}
To completely remove all information about the events that were not detected, we must also marginalize over the number of events that are not detected ($N \geq N_\mathbb{D}$).
This produces a term like
\begin{align}
    \sum\limits_{N=N_\mathbb{D}}^\infty \frac{1}{(N-N_\mathbb{D})!} \left[ N_E P(\cancel{\mathbb{D}}|\Lambda) \right]^{N-N_\mathbb{D}}
        & = e^{N_E P(\cancel{\mathbb{D}}|\Lambda)} \nonumber \\
        & = e^{N_E (1 - P(\mathbb{D}|\Lambda))}
\end{align}
Simplifying yields
\begin{align}
    p(\{\theta, x, \mathbb{D}, \cancel{\mathbb{F}}\}_i^{N_\mathbb{D}-N_\mathbb{F}}, \{\theta, x, f, & \mathbb{D}, \mathbb{F}\}_j^{N_\mathbb{F}}, N_\mathbb{F}, N_\mathbb{D} | N_E, \Lambda) \nonumber \\
    & = N_E^{N_\mathbb{D}} e^{-N_E P(\mathbb{D}|\Lambda)} \frac{1}{(N_\mathbb{D}-N_\mathbb{F})!} \left[ \prod\limits_{i=1}^{N_\mathbb{D}-N_\mathbb{F}} P(\cancel{\mathbb{F}}|\mathbb{D}, x_{i}) P(\mathbb{D}|x_{i}) p(x_{i}|\theta_i) p(\theta_i|\Lambda) \right] \nonumber \\
    & \quad \quad \quad \quad \times \frac{1}{N_\mathbb{F}!} \left[ \prod\limits_{j=1}^{N_\mathbb{F}} P(\mathbb{F}|\mathbb{D}, x_{j}) P(\mathbb{D}|x_{j}) p(x_{j}|\theta_j) p(f_{j}|\theta_j) p(\theta_j|\Lambda) \right] \label{eq:top-down joint}
\end{align}
which is identical to Eq.~\ref{eq:main result joint}.


\subsection{Bottom-Up Derivation}
\label{sec:bottom-up}

We now start with the distribution of observed data (conditioned on detection) and work backwards.
The likelihood of the data for an individual system ($x$) conditioned on detection ($\mathbb{D}$) and the true parameters ($\theta$) is
\begin{equation}
    p(x|\mathbb{D}, \theta) = \frac{P(\mathbb{D}|x) p(x|\theta)}{P(\mathbb{D}|\theta)}
\end{equation}
In order to consider the joint distribution of both the data and the true parameters, we also need to use the conditioned prior
\begin{equation}
    p(\theta|\mathbb{D},\Lambda) = \frac{P(\mathbb{D}|\theta) p(\theta|\Lambda)}{P(\mathbb{D}|\Lambda)}
\end{equation}
Combining them yields
\begin{align}
    p(x,\theta|\mathbb{D},\Lambda)
        & = p(x|\theta, \mathbb{D}) p(\theta|\mathbb{D}, \Lambda) \nonumber \\
        & = \frac{P(\mathbb{D}|x)}{P(\mathbb{D}|\Lambda)} p(x|\theta) p(\theta|\Lambda)
\end{align}
where the factors of $P(\mathbb{D}|\theta)$ cancel.

Now, consider two separate sets of events (those that are detected and followed-up vs. those that are detected but not followed-up).
We repeat this logic and obtain
\begin{align}
    p(x,\theta|\mathbb{D},\cancel{\mathbb{F}},\Lambda)
        & = \frac{P(\mathbb{D}, \cancel{\mathbb{F}}|x)}{P(\mathbb{D},\cancel{\mathbb{F}}|\Lambda)} p(x|\theta) p(\theta|\Lambda) \\
    p(x, f, \theta|\mathbb{D},\mathbb{F},\Lambda)
        & = \frac{P(\mathbb{D}, \mathbb{F}|x)}{P(\mathbb{D},\mathbb{F}|\Lambda)} p(x|\theta) p(f|\theta) p(\theta|\Lambda)
\end{align}
Note that we assume $p(x, f|\theta) = p(x|\theta) p(f|\theta)$ and that $f$ only depends on $\theta$ (i.e., does not depend on $\mathbb{D}$ or $\mathbb{F}$ directly), as shown in Fig.~\ref{fig:dag}.

If we have $N_\mathbb{D}-N_\mathbb{F}$ i.i.d. systems that were detected but not followed-up and $N_\mathbb{F}$ i.i.d. systems that were detected and followed-up, we multiply these distributions to obtain
\begin{align}
    p(\{x, \theta\}_i^{N_\mathbb{D}-N_\mathbb{F}}, \{x,f,\theta\}_j^{N_\mathbb{F}} & | \{\mathbb{D},\cancel{\mathbb{F}}\}_i^{N_\mathbb{D}-N_\mathbb{F}}, \{\mathbb{D}, \mathbb{F}\}_j^{N_\mathbb{F}}, N_\mathbb{D}, N_\mathbb{F}, \Lambda) \nonumber \\
        & = \left[ \prod\limits_{i=1}^{N_\mathbb{D}-N_\mathbb{F}} \frac{P(\mathbb{D}, \cancel{\mathbb{F}}|x_{i})}{P(\mathbb{D},\cancel{\mathbb{F}}|\Lambda)} p(x_{i}|\theta_{i}) p(\theta_{i}|\Lambda) \right] \left[ \prod\limits_{j=1}^{N_\mathbb{F}} \frac{P(\mathbb{D}, \mathbb{F}|x_{j})}{P(\mathbb{D},\mathbb{F}|\Lambda)} p(x_{j}|\theta_j) p(f_{j}|\theta_j) p(\theta_j|\Lambda) \right] \label{eq:bottom-up joint|DF}
\end{align}
Eq.~\ref{eq:bottom-up joint|DF} is essentially the same as Eq.~\ref{eq:single-event|N}.

We additionally include the uncertainty on $N_\mathbb{D}-N_\mathbb{F}$ and $N_\mathbb{F}$, which are Poisson distributed with means determined by the expected number of astrophysical systems $N_E$ and the corresponding detection probabilities.
\begin{align}
    p(N_\mathbb{D}-N_\mathbb{F}|N_E,\Lambda) & = \frac{1}{(N_\mathbb{D}-N_\mathbb{F})!} \left[ N_E P(\mathbb{D}, \cancel{\mathbb{F}}|\Lambda) \right]^{N_\mathbb{D}-N_\mathbb{F}} e^{-N_E P(\mathbb{D},\cancel{\mathbb{F}}|\Lambda)} \\
    p(N_\mathbb{F}|N_E,\Lambda) & = \frac{1}{N_\mathbb{F}!} \left[ N_E P(\mathbb{D}, \mathbb{F}|\Lambda) \right]^{N_\mathbb{F}} e^{-N_E P(\mathbb{D},\mathbb{F}|\Lambda)}
\end{align}
Note that conditioning on the number of events in each set ($N_\mathbb{D}-N_\mathbb{F}$ and $N_\mathbb{F}$, respectively) is equivalent to conditioning on the set of detection indicators ($\{\mathbb{D}, \cancel{\mathbb{F}}\}$ and $\{\mathbb{D}, \mathbb{F}\}$, respectively).
Therefore, when we combine the distribution conditioned on ($N_\mathbb{D}, N_\mathbb{F}, \{\mathbb{D}, \cancel{\mathbb{F}}\}_i^{N_\mathbb{D}-N_\mathbb{F}}, \{\mathbb{D}, \mathbb{F}\}_j^{N_\mathbb{F}}$) with the Poisson distributions for ($N_\mathbb{D}-N_\mathbb{F}, N_\mathbb{F} | N_E, \Lambda$), we should move all of ($N_\mathbb{D}, N_\mathbb{F}, \{\mathbb{D}, \cancel{\mathbb{F}}\}_i^{N_\mathbb{D}-N_\mathbb{F}}, \{\mathbb{D}, \mathbb{F}\}_j^{N_\mathbb{F}}$) to the left-hand-side of the conditional bar.
This yields
\begin{align}
p(\{\mathbb{D}, \cancel{\mathbb{F}}, x, & \theta\}_i^{N_\mathbb{D}-N_\mathbb{F}}, \{\mathbb{D}, \mathbb{F}, x, f, \theta\}_j^{N_\mathbb{F}}, N_\mathbb{D}, N_\mathbb{F} | N_E, \Lambda) \nonumber \\
        & = \frac{1}{(N_\mathbb{D}-N_\mathbb{F})!} \left[ N_E P(\mathbb{D}, \cancel{\mathbb{F}}|\Lambda) \right]^{N_\mathbb{D}-N_\mathbb{F}} e^{-N_E P(\mathbb{D},\cancel{\mathbb{F}}|\Lambda)} \left[ \frac{1}{P(\mathbb{D},\cancel{\mathbb{F}}|\Lambda)^{N_\mathbb{D}-N_\mathbb{F}}} \prod\limits_{i=1}^{N_\mathbb{D}-N_\mathbb{F}} P(\mathbb{D}, \cancel{\mathbb{F}}|x_{i}) p(x_{i}|\theta_{i}) p(\theta_{i}|\Lambda) \right] \nonumber \\
        & \quad \quad \quad \quad \times \frac{1}{N_\mathbb{F}!} \left[ N_E P(\mathbb{D}, \mathbb{F}|\Lambda) \right]^{N_\mathbb{F}} e^{-N_E P(\mathbb{D},\mathbb{F}|\Lambda)} \left[ \frac{1}{P(\mathbb{D},\mathbb{F}|\Lambda)^{N_\mathbb{F}}} \prod\limits_{j=1}^{N_\mathbb{F}} P(\mathbb{D}, \mathbb{F}|x_{j}) p(x_{j}|\theta_j) p(f_{j}|\theta_j) p(\theta_j|\Lambda) \right] \nonumber \\
        & = \frac{1}{N_F!(N_\mathbb{D}-N_\mathbb{F})!} N_E^{N_\mathbb{D}} e^{-N_E [P(\mathbb{D},\cancel{\mathbb{F}}|\Lambda) + P(\mathbb{D},\mathbb{F}|\Lambda)]} \nonumber \\
        & \quad \quad \quad \quad \times \left[ \prod\limits_{i=1}^{N_\mathbb{D}-N_\mathbb{F}} P(\mathbb{D}, \cancel{\mathbb{F}}|x_{i}) p(x_{i}|\theta_{i}) p(\theta_{i}|\Lambda) \right] \left[ \prod\limits_{j=1}^{N_\mathbb{F}} P(\mathbb{D}, \mathbb{F}|x_{j}) p(x_{j}|\theta_j) p(f_{j}|\theta_j) p(\theta_j|\Lambda) \right] \nonumber \\
        & = \frac{1}{N_\mathbb{F}!(N_\mathbb{D}-N_\mathbb{F})!} N_E^{N_\mathbb{D}} e^{-N_E P(\mathbb{D}|\Lambda)} \left[ \prod\limits_{i=1}^{N_\mathbb{D}-N_\mathbb{F}} P(\mathbb{D}, \cancel{\mathbb{F}}|x_{i}) p(x_{i}|\theta_{i}) p(\theta_{i}|\Lambda) \right] \left[ \prod\limits_{j=1}^{N_\mathbb{F}} P(\mathbb{D}, \mathbb{F}|x_{j}) p(x_{j}|\theta_j) p(f_{j}|\theta_j) p(\theta_j|\Lambda) \right] \label{eq:bottom-up joint}
\end{align}
where in the last line we recognize that the sum in the exponent of the combined Poisson term is just the marginalization over whether events are followed-up or not: $P(\mathbb{D}|\Lambda) = P(\mathbb{D}, \cancel{\mathbb{F}}|\Lambda) + P(\mathbb{D},\mathbb{F}|\Lambda)$.
Eq.~\ref{eq:bottom-up joint} is identical to Eq.~\ref{eq:main result joint} and~\ref{eq:top-down joint}.


\section{Additional Robustness checks}
\label{sec:robustness checks}

\begin{figure}
    \begin{center}
        \includegraphics[width=0.49\textwidth]{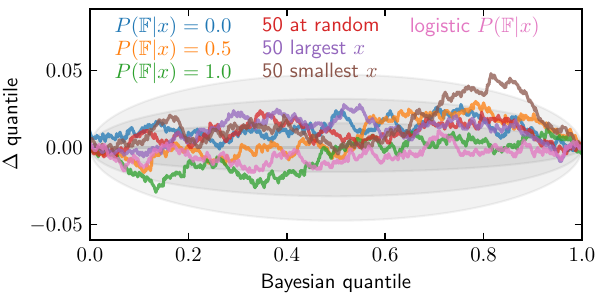}
        \includegraphics[width=0.49\textwidth]{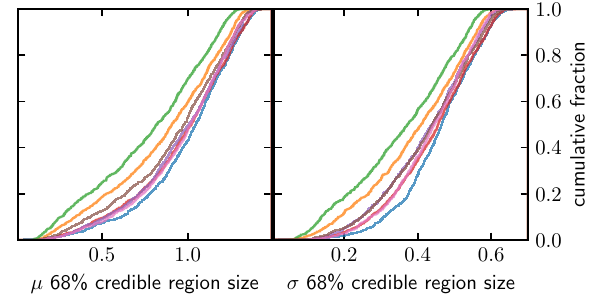}
    \end{center}
    \caption{
        (\emph{left, colored lines}) Differences between Bayesian and frequentist quantiles ($\Delta$ quantile) vs. Bayesian quantile (corresponding to the highest-probability-density region that marginally contains the true parameters) for posteriors inferred using different follow-up strategies and (\emph{grey shading}) expected uncertainty from the finite number of trials used.
        (\emph{right}) Cumulative fractions of trials as a function of the width of the \result{68\% symmetric posterior credible region} for the population mean ($\mu$) and standard deviation ($\sigma$).
        The widths of the corresponding prior credible regions are \result{2.0} and \result{2.7}, respectively.
        We generated \result{1000} mock catalogs and considered the following follow-up strategies: (\emph{blue}) follow-up nothing, (\emph{orange}) follow-up events randomly with a fixed probability of 50\%, (\emph{green}) follow-up everything, (\emph{red}) follow-up exactly 50 events chosen at random, (\emph{purple}) follow-up the 50 events with the largest $x$, (\emph{brown}) follow-up the 50 events with the smallest $x$, and (\emph{pink}) follow-up events with a logistic probability (Eq.~\ref{eq:simplified toy model follow up}) with \result{$X_\mathbb{F}=5$ and $S_\mathbb{F}=1$}.
        Comparing these to a uniform distribution yields KS $p$-values of \result{0.41}, \result{0.32}, \result{0.33}, \result{0.60}, \result{0.39}, \result{0.02}, and \result{0.93}, respectively.
    }
    \label{fig:simplified toy model coverage}
\end{figure}

We consider the simple Gaussian model from Sec.~\ref{sec:gaussian toy model} in more detail to confirm the derivations in Appendix~\ref{sec:derivations}.
Using this model, we generate \result{1000} catalogs.
Each contains \result{500} detected events.
\result{Each catalog also has $\mu$ and $\sigma$ separately drawn from a standard unit-normal distribution and a uniform distribution between 1 and 5, respectively.}
\result{We assume $\sigma_x=1$, $\sigma_f=0.1$, $X_\mathbb{D}=0$, and $S_\mathbb{D}=0.1$.}
Using the same catalogs, but applying different follow-up strategies, we then sample from the posterior: $p(\mu,\sigma|\{x,\mathbb{D},\cancel{\mathbb{F}}\}^{N_\mathbb{D}-N_\mathbb{F}}, \{x,f,\mathbb{D},\mathbb{F}\}^{N_\mathbb{F}}, N_\mathbb{D}, N_\mathbb{F})$.
Importantly, we use a prior on $(\mu, \sigma)$ that matches the true distributions from which catalogs were drawn.

As discussed in Sec.~\ref{sec:gaussian toy model}, Fig.~\ref{fig:simplified toy model} shows an example of one such catalog and the corresponding posterior.
Fig.~\ref{fig:simplified toy model coverage} shows coverage tests and distributions of the widths of \result{68\%} symmetric posterior credible regions for \result{7} different follow-up strategies.
Proper coverage corresponds to a horizontal line at zero on the left panel of Fig.~\ref{fig:simplified toy model coverage}, and the shaded bands approximate 1, 2, and 3-$\sigma$ fluctuations that could be due to the finite number of trials performed.
Thus, Fig.~\ref{fig:simplified toy model coverage} demonstrates that proper coverage is obtained by an inference that never explicitly models the follow-up selection probability when the follow-up decision corresponds to either
\begin{enumerate*}[label=(\roman*)]
    \item follow-up nothing (i.e., $P(\mathbb{F}) = 0$),
    \item follow-up randomly with $P(\mathbb{F}) = 1/2$ (i.e., a variable number of events are followed-up),
    \item follow-up everything (i.e., $P(\mathbb{F}) = 1$),
    \item follow-up 50 events and select them at random (i.e., a fixed number of events are followed-up),
    \item follow-up the 50 events with the largest $x$,
    \item follow-up the 50 with the smallest $x$, or
    \item follow-up with a logistic probability (Eq.~\ref{eq:simplified toy model follow up}) with $X_\mathbb{F}=5$ and $S_\mathbb{F}=1$, which is also shown in Fig.~\ref{fig:simplified toy model}.
\end{enumerate*}
In all cases, we obtain informative posteriors (compared to the prior) with correct coverage.
The right panel of Fig.~\ref{fig:simplified toy model coverage} demonstrates that the reason for obtaining correct coverage is not simply because the posterior equals the prior.
Rather, the population posterior is always tighter than its prior.
It also shows that posteriors that include follow-up for all events  systematically have the smallest posterior credible regions, and posteriors that include no follow-up systematically have the widest posterior credible regions. 


\section{More complicated follow-up decision processes and active learning}
\label{sec:active learning}

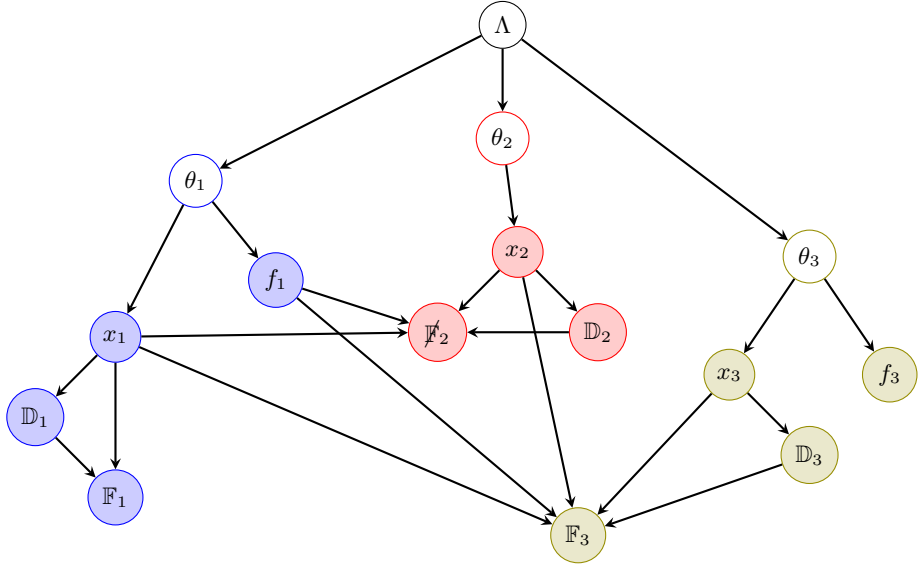
\begin{figure}[b]
    \begin{center}
        \begin{tikzpicture}[node distance=1.5cm]
        \tikzstyle {node} = [circle, text centered, draw=black, fill=white] ;
        \tikzstyle {node1} = [circle, text centered, draw=blue, fill=white] ;
        \tikzstyle {recordednode1} = [circle, text centered, draw=blue, fill=blue, fill opacity=0.2, text opacity=1.0] ;
        \tikzstyle {node2} = [circle, text centered, draw=red, fill=white] ;
        \tikzstyle {recordednode2} = [circle, text centered, draw=red, fill=red, fill opacity=0.2, text opacity=1.0] ;
        \tikzstyle {node3} = [circle, text centered, draw=olive, fill=white] ;
        \tikzstyle {recordednode3} = [circle, text centered, draw=olive, fill=olive, fill opacity=0.2, text opacity=1.0] ;
        \tikzstyle {arrow} = [thick, ->, >=stealth, draw=black] ;
        \node (L) [node] {$\Lambda$} ;
        \node (t1) [node1, below left of=L, xshift=-3.0cm, yshift=-1.0cm] {$\theta_1$} ;
        \node (x1) [recordednode1, below left of=t1, yshift=-1.0cm] {$x_1$} ;
        \node (f1) [recordednode1, below right of=t1, yshift=-0.25cm] {$f_1$} ;
        \node (D1) [recordednode1, below left of=x1] {$\mathbb{D}_1$} ;
        \node (F1) [recordednode1, below right of=D1] {$\mathbb{F}_1$} ;
        \node (t2) [node2, below of=L, xshift=+0.0cm, yshift=-0.0cm] {$\theta_2$} ;
        \node (x2) [recordednode2, below of=t2, xshift=+0.2cm, yshift=-0.0cm] {$x_2$} ;
        \node (D2) [recordednode2, below right of=x2] {$\mathbb{D}_2$} ;
        \node (notF2) [recordednode2, below left of=x2] {$\cancel{\mathbb{F}}_2$} ;
        \node (t3) [node3, below right of=L, xshift=+3.0cm, yshift=-2.0cm] {$\theta_3$} ;
        \node (x3) [recordednode3, below left of=t3, yshift=-0.5cm] {$x_3$} ;
        \node (f3) [recordednode3, below right of=t3, yshift=-0.5cm] {$f_3$} ;
        \node (D3) [recordednode3, below right of=x3] {$\mathbb{D}_3$} ;
        \node (F3) [recordednode3, below left of=D3, xshift=-2.0cm] {$\mathbb{F}_3$} ;
        \draw [arrow] (L) -- (t1) ;
        \draw [arrow] (t1) -- (x1) ;
        \draw [arrow] (t1) -- (f1) ;
        \draw [arrow] (x1) -- (D1) ;
        \draw [arrow] (x1) -- (F1) ;
        \draw [arrow] (D1) -- (F1) ;
        \draw [arrow] (L) -- (t2) ;
        \draw [arrow] (t2) -- (x2) ;
        \draw [arrow] (x2) -- (D2) ;
        \draw [arrow] (x1) -- (notF2) ;
        \draw [arrow] (f1) -- (notF2) ;
        \draw [arrow] (x2) -- (notF2) ;
        \draw [arrow] (D2) -- (notF2) ;
        \draw [arrow] (L) -- (t3) ;
        \draw [arrow] (t3) -- (x3) ;
        \draw [arrow] (t3) -- (f3) ;
        \draw [arrow] (x3) -- (D3) ;
        \draw [arrow] (x1) -- (F3) ;
        \draw [arrow] (f1) -- (F3) ;
        \draw [arrow] (x2) -- (F3) ;
        \draw [arrow] (x3) -- (F3) ;
        \draw [arrow] (D3) -- (F3) ;
        \end{tikzpicture}
    \end{center}
    \caption{
        A simple model of active learning in which future follow-up decisions are based on all previous data through, presumably, an inference of the population constructed with all previously recorded data.
        Nodes are colored according to which events they are associated with (\emph{events 1, 2, and 3 are colored blue, red, and green, respectively}).
    }
    \label{fig:active learning}
\end{figure}

Consider the model in Fig.~\ref{fig:active learning} in which there are three events and $\mathbb{F}_3$ depends directly on all previously observed data, including follow-up observations: ($x_2, x_1, f_1$).
This will produce a joint distribution like
\begin{align}
    p(\{\theta, x, \mathbb{D}\}_{i=1,2,3}, \{f, \mathbb{F}\}_{j=1,3}, \cancel{\mathbb{F}}_2 | \Lambda)
        & \propto P(\mathbb{F}_3|\mathbb{D}_3,x_3,x_2,x_1,f_1) P(\mathbb{D}_3|x_3)p(x_3|\theta_3)p(f_3|\theta_3)p(\theta_3|\Lambda) \nonumber \\
        & \quad \quad \quad \quad \times P(\cancel{\mathbb{F}}_2|\mathbb{D}_2,x_2,x_1,f_1)P(\mathbb{D}_2|x_2)p(x_2|\theta_2)p(\theta_2|\Lambda) \nonumber \\
        & \quad \quad \quad \quad \quad \quad \quad \quad \times P(\mathbb{F}_1|\mathbb{D}_1,x_1)P(\mathbb{D}_1|x_1)p(x_1|\theta_1)p(f_1|\theta_1)p(\theta_1|\Lambda)
\end{align}
Just as in constructing Eq.~\ref{eq:main result post} from Eq.~\ref{eq:main result joint}, all terms involving $\mathbb{F}$ will cancel when we construct a posterior conditioned on $\{x, f, \mathbb{D}, \mathbb{F}\}$.
This means that we can use all previously obtained data when determining whether to follow-up new events (for example, by inferring the underlying population and only investing in follow-up for events with uncertain classifications) without explicitly modeling the follow-up decision process as long as we condition on all the data that went into that decision process.

Similarly, $\mathbb{F}_i$ can depend on the entire set of the original catalog data $\{x\}_i^{N_\mathbb{D}}$ rather than just on the $x_i$ that corresponds to the event being followed-up (Fig.~\ref{fig:dag}).


\section{Discarding Follow-up Data}
\label{sec:discarding follow-up}

For the sake of completeness, we now show that one cannot discard follow-up data ($f$) based on the same follow-up data without introducing an additional selection that must be modeled explicitly within the posterior~\citep[see also][]{Essick:2022}.
This is true even though one can discard follow-up observations based on the original catalog data ($x$) without modeling that selection.

We introduce another indicator for whether follow-up data is retained ($\mathbb{R}$) or discarded ($\cancel{\mathbb{R}}$) based on $f$.
Starting from Eq.~\ref{eq:main result joint}, we obtain a joint distribution
\begin{align}
    p(\{\theta,x,\mathbb{D},\cancel{\mathbb{F}}\}_i^{N_\mathbb{D}-N_\mathbb{F}}, & \{\theta,x,f,\mathbb{D},\mathbb{F},\cancel{\mathbb{R}}\}_j^{N_\mathbb{F}-N_\mathbb{R}}, \{\theta,x,f,\mathbb{D},\mathbb{F},\mathbb{R}\}_k^{N_\mathbb{R}}, N_\mathbb{D}, N_\mathbb{F}, N_\mathbb{R} | N_E, \Lambda) \nonumber \\
        & = N_E^{N_\mathbb{D}} e^{-N_E P(\mathbb{D}|\Lambda)} \frac{1}{(N_\mathbb{D}-N_\mathbb{F})!} \prod\limits_i^{N_\mathbb{D}-N_\mathbb{F}} P(\mathbb{D}, \cancel{\mathbb{F}}|x_i) p(x_i|\theta) p(\theta|\Lambda) \nonumber \\
        & \quad \quad \quad \quad \times \frac{1}{(N_\mathbb{F}-N_\mathbb{R})!} \prod\limits_j^{N_\mathbb{F}-N_\mathbb{R}} P(\cancel{\mathbb{R}}|\mathbb{D}, \mathbb{F}, x_j, f_j) P(\mathbb{D},\mathbb{F}|x_j) p(f_j|\theta) p(x_j|\theta) p(\theta | \Lambda) \nonumber \\
        & \quad \quad \quad \quad \quad \quad \quad \quad \times \frac{1}{N_\mathbb{R}!} \prod\limits_k^{N_\mathbb{R}}  P(\mathbb{R}|\mathbb{D}, \mathbb{F}, x_k, f_k) P(\mathbb{D},\mathbb{F}|x_k) p(f_k|\theta) p(x_k|\theta) p(\theta|\Lambda)
\end{align}
where only $N_\mathbb{R} \leq N_\mathbb{F}$ events with follow-up are retained.
If we now marginalize over $f_j$ for the $N_\mathbb{F}-N_\mathbb{R}$ events with $\cancel{\mathbb{R}}$, we obtain terms like
\begin{equation}
    \int df_j P(\cancel{\mathbb{R}}|\mathbb{D},\mathbb{F}, x_j, f_j) p(f_j|\theta_j) = P(\cancel{\mathbb{R}}|\mathbb{D},\mathbb{F}, x_j, \theta_j)
\end{equation}
Therefore, when we compute the posterior, these terms will not factor out of the integrals over $\theta_j$ and will not cancel.
If one attempts to additionally integrate over $\theta_j$, these terms will depend on $\Lambda$ and still will not cancel.
The only way for the cancellation to persist is if $(\mathbb{F}, \mathbb{R} \perp f | \mathbb{D}, x)$, i.e., if both the decisions to follow-up and/or retain the follow-up data are based only on the original catalog's data.
Thus, we reiterate that all follow-up data must be included in population inferences that do not modeled follow-up decisions, even if some followed-up events end up being contaminants or the follow-up data is unconstraining.
In this case, a mixture model of contaminants and target systems may be needed, as described in Sec.~\ref{sec:rare systems} and Appendix~\ref{sec:gaia model}.


\section{Related DAGs and modeling the ``detected distribution''}
\label{sec:related dag}

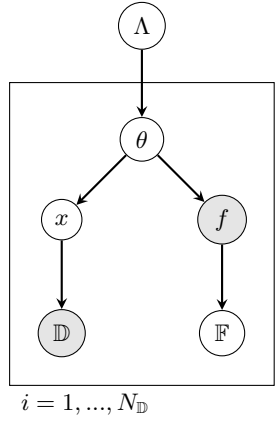
\begin{figure*}
    \begin{center}
    \begin{tikzpicture}[node distance=1.5cm]
        \tikzstyle {node} = [circle, text centered, draw=black, fill=white];
        \tikzstyle {recordednode} = [circle, text centered, draw=black, fill=gray, fill opacity=0.2, text opacity=1.0];
        \tikzstyle {arrow} = [thick, ->, >=stealth, draw=black]
        \node (L) [node] {$\Lambda$} ;
        \draw [draw=black] (-1.75, -0.75) rectangle (+1.75, -4.75);
        \node at (-0.75, -5.00) {$i=1,...,N_\mathbb{D}$};
        \node (t) [node, below of=L] {$\theta$} ;
        \node (x) [node, below left of=t] {$x$} ;
        \node (f) [recordednode, below right of=t] {$f$} ;
        \node (D) [recordednode, below of=x] {$\mathbb{D}$} ;
        \node (F) [node, below of=f] {$\mathbb{F}$} ;
        %
        %
        \draw [arrow] (L) -- (t) ;
        \draw [arrow] (t) -- (x) ;
        \draw [arrow] (t) -- (f) ;
        \draw [arrow] (x) -- (D) ;
        \draw [arrow] (f) -- (F) ;
    \end{tikzpicture}
    \end{center}
    \caption{
        DAG representing a related observational scheme in which events are observed in two instruments with data $x$ and $f$, respectively.
        We select events based on $x$ ($\mathbb{D}$) but do not condition on $x$ itself.
        Eq.~\ref{eq:related dag joint} assumes that we do not select based on $f$ ($\mathbb{F}$) whereas Eq.~\ref{eq:related dag messy joint} does.
    }
    \label{fig:related dag}
\end{figure*}

We now consider a related DAG (Fig.~\ref{fig:related dag}) that describes an inference based on follow-up data $f$ when we select events based on some other dataset $x$ but do not condition on $x$ directly.

Start from the joint distribution for $(\{x,f,\mathbb{D}\},N_\mathbb{D}|N_E, \Lambda)$
\begin{equation}
    p(\{\theta, x, f, \mathbb{D}\}_i^{N_\mathbb{D}}, N_\mathbb{D}| N_E, \Lambda) = \frac{1}{N_\mathbb{D}!} N_E^{N_\mathbb{D}} e^{-N_E P(\mathbb{D}|\Lambda)} \prod\limits_{i=1}^{N_\mathbb{D}} P(\mathbb{D}|x_i) p(x_i|\theta_i) p(f_i|\theta_i) p(\theta_i|\Lambda)
\end{equation}
and marginalize over $x$ to obtain
\begin{align}
    p(\{\theta, f, \mathbb{D}\}_i^{N_\mathbb{D}}, N_\mathbb{D}| N_E, \Lambda) 
        & = \frac{1}{N_\mathbb{D}!} N_E^{N_\mathbb{D}} e^{-N_E P(\mathbb{D}|\Lambda)} \prod\limits_{i=1}^{N_\mathbb{D}} p(f_i|\theta_i) P(\mathbb{D}|\theta_i) p(\theta_i|\Lambda) \nonumber \\
        & = \frac{1}{N_\mathbb{D}!} N_E^{N_\mathbb{D}} e^{-N_E P(\mathbb{D}|\Lambda)} \prod\limits_{i=1}^{N_\mathbb{D}} p(f_i|\theta_i)  p(\theta_i|\mathbb{D},\Lambda) P(\mathbb{D}|\Lambda) 
\end{align}
where we used the identity $P(\mathbb{D}|\theta)p(\theta|\Lambda) = p(\theta|\mathbb{D},\Lambda)P(\mathbb{D}|\Lambda)$ in the second line.
If we additionally change variables from the expected number of astrophysical systems ($N_E$) to the expected number of detected systems ($K_E = N_E P(\mathbb{D}|\Lambda)$),
we obtain
\begin{align}
p(\{\theta, f, \mathbb{D}\}_i^{N_\mathbb{D}}, N_\mathbb{D}| K_E, \Lambda) 
        & = \frac{1}{N_\mathbb{D}!} \left(\frac{K_E}{P(\mathbb{D}|\Lambda)}\right)^{N_\mathbb{D}} e^{-K_E} \prod\limits_{i=1}^{N_\mathbb{D}} p(f_i|\theta_i) P(\theta_i|\mathbb{D},\Lambda) P(\mathbb{D}|\Lambda) \nonumber \\
        & = \frac{1}{N_\mathbb{D}!} K_E^{N_\mathbb{D}} e^{-K_E} \prod\limits_{i=1}^{N_\mathbb{D}} p(f_i|\theta_i)  p(\theta_i|\mathbb{D},\Lambda) \label{eq:related dag joint}
\end{align}
Eq.~\ref{eq:related dag joint} is exactly what is predicted by the ``unphysical DAG'' from~\citet{Essick:2024}.
This why the unphysical DAG is said to erroneously consider independent noise realizations for detection ($x$) and parameter estimation ($f$), which is incorrect when detection and parameter estimation are performed with the same data.
However, this does mean that one can ``fit the detected distribution,'' for example, with events that are detected as gamma-ray bursts (GRBs) and then followed-up with GW parameter estimation (often called a ``triggered search''in the GW community).\footnote{Other authors have shown that it is also possible, in principle, to ``fit the detected distribution'' in the standard hierarchical inference~\citep{Toubiana:2025}, although the likelihood is different than Eq.~\ref{eq:related dag joint}. Instead of modeling $P(\mathbb{D}|\Lambda)$, one must precisely model $P(\mathbb{D}|\theta) \ \forall \ \theta$, which may be computationally prohibitive and/or prone to poorly-quantified systematic uncertainties.}

Note, however, that this is somewhat spoiled if we also select events based on $f$ (i.e., $\mathbb{F}$ is observed in Fig.~\ref{fig:related dag}).
In this case, the joint distribution is
\begin{equation}
    p(\{\theta, x,f,\mathbb{D},\mathbb{F}\}_j^{N_\mathbb{F}} | N_E, \Lambda) = \frac{1}{N_\mathbb{F}!} N_E^{N_\mathbb{F}} e^{-N_E P(\mathbb{D}, \mathbb{F}|\Lambda)} \prod\limits_{j=1}^{N_\mathbb{F}} P(\mathbb{D}|x_j) p(x_j|\theta_j) P(\mathbb{F}|f_j) p(f_j|\theta_j) p(\theta_j|\Lambda)
\end{equation}
Marginalizing over $x$ and changing variables from $N_E$ to $K_E = N_E P(\mathbb{D}, \mathbb{F}|\Lambda)$ yields
\begin{equation} 
    p(\{\theta, f,\mathbb{D},\mathbb{F}\}_j^{N_\mathbb{F}} | K_E, \Lambda) = \frac{1}{N_\mathbb{F}!} K_E^{N_\mathbb{F}} e^{-K_E} \prod\limits_{j=1}^{N_\mathbb{F}} \frac{P(\mathbb{F}|f_j)}{P(\mathbb{F}|\mathbb{D},\Lambda)} p(f_j|\theta_j) p(\theta_j|\mathbb{D}, \Lambda) \label{eq:related dag messy joint}
\end{equation}
In this case, as in Appendix~\ref{sec:discarding follow-up}, the terms $P(\mathbb{F}|\mathbb{D},\Lambda)$ will not cancel when we condition Eq.~\ref{eq:related dag messy joint} on ($f,\mathbb{D},\mathbb{F}$) to obtain a posterior, and we do not obtain the same simple ``unphysical DAG'' result (Eq.~\ref{eq:related dag joint}).
Instead, we obtain
\begin{equation}
    p(\{\theta\}_j^{N_\mathbb{F}}, K_E, \Lambda|\{f,\mathbb{D},\mathbb{F}\}_j^{N_\mathbb{F}}) \propto K_E^{N_\mathbb{F}} e^{-K_E} \left(\frac{1}{P(\mathbb{F}|\mathbb{D},\Lambda)}\right)^{N_\mathbb{F}} \prod\limits_{j=1}^{N_\mathbb{F}} p(f_j|\theta_j) p(\theta_j|\mathbb{D},\Lambda)
\end{equation}
and one must explicitly model $P(\mathbb{F}|\mathbb{D},\Lambda)$ in addition to $p(\theta|\mathbb{D},\Lambda)$.
Specifically,
\begin{align}
    P(\mathbb{F}|\mathbb{D},\Lambda) 
        & = \frac{\int d\theta dx df \, P(\mathbb{F}|f) p(f|\theta) P(\mathbb{D}|x) p(x|\theta) p(\theta|\Lambda)}{P(\mathbb{D}|\Lambda)} \nonumber \\
        & = \int d\theta df \, P(\mathbb{F}|f) p(f|\theta) \frac{P(\mathbb{D}|\theta)p(\theta|\Lambda)}{P(\mathbb{D}|\Lambda)} \nonumber \\
        & = \int d\theta df \, P(\mathbb{F}|f) p(f|\theta) p(\theta|\mathbb{D},\Lambda)
\end{align}
This is similar to the ``physical DAG'' described in~\citet{Essick:2024} in which one must model $P(\mathbb{D}|\Lambda)$ and $p(\theta|\Lambda)$ simultaneously.
What's more, if $\mathbb{D}$ only depends on nuisance parameters, like the orbital inclination of compact binary systems, but does not depend other parameters of interest, like cosmological parameters, then one can simply marginalize over the nuisance parameters while simultaneously fitting for $p(\theta|\mathbb{D},\Lambda)$ and extract an unbiased estimate for the (cosmological) parameters of interest.
This is similar to the proposal in~\citet{Salvarese:2024}.

Note also that~\citet{Chen:2023} propose an inference that includes all events detected by GWs, modifies the likelihood to include additional EM data, and fits for the unknown selection effect explicitly.
A similar approach is proposed in~\citet{Mancarella:2024} (only modeling events detected in both GWs and EM), and~\citet{Mould:2023} considers a somewhat simpler scenario.
This approach is related to, but distinct from, our main result.
They assume that $\mathbb{F}$ (successful follow-up and/or EM counterpart detection) depends on $\theta$ through $f$ (i.e., $\mathbb{F} \perp \theta \ | \ f$) according to Fig.~\ref{fig:related dag}, whereas our main result assumes that $\mathbb{F} \perp f \ | \ \theta$ (Fig.~\ref{fig:dag}).
In that sense,~\citet{Chen:2023} and~\citet{Mancarella:2024} propose set-ups closer to what we discuss in Appendix~\ref{sec:discarding follow-up}, which is why they must model the follow-up selection process.


\section{Bright Sirens model}
\label{sec:bright sirens model}

Below, we describe the model used to construct Fig.~\ref{fig:bright sirens} in Sec.~\ref{sec:gw cosmology}, which is similar to the model explored in~\citet{Farah:2024}.
We assume a uniform distribution in source-frame chirp masses ($\mathcal{M}_\mathrm{src}$) and a uniform-in-volume distribution over distance ($D$).
The redshift ($z$) is assumed to scale linearly with $D$ according to the Hubble relation (Eq.~\ref{eq:Hubble}), and the detector-frame mass is the redshifted version of the source-frame mass (Eq.~\ref{eq:mdet from msrc}).
The optimal signal-to-noise ratio ($\rho$) of each source approximately follows the scaling expected for an inspiral-dominated post-Newtonian waveform truncated at the inner-most stable circular orbit (ISCO).
Observed properties ($\hat{\rho}$, $\hat{\mathcal{M}}_\mathrm{det}$, and $\hat{z}$) are normally distributed around the corresponding true values.
We additionally select events with a simple threshold on $\hat{\rho}$.

\begin{align}
    p(\mathcal{M}_\mathrm{src}|\Lambda)
        & = \frac{1}{\mathcal{M}_\mathrm{max} - \mathcal{M}_\mathrm{min}} \Theta(\mathcal{M}_\mathrm{min} \leq \mathcal{M}_\mathrm{src} \leq \mathcal{M}_\mathrm{max}) \\
    p(D|\Lambda)
        & = \frac{3D^2}{D_\mathrm{max}^3} \Theta(0 \leq D \leq D_\mathrm{max}) \\
    z
        & = \frac{H_0}{c} D \label{eq:Hubble} \\
    \mathcal{M}_\mathrm{det}
        & = \mathcal{M}_\mathrm{src} (1+z) \label{eq:mdet from msrc} \\
    \rho
        & = \frac{\mathcal{M}_\mathrm{det}^{5/6}}{D}\left( 1 - \left(\frac{f_\mathrm{isco}}{f_\mathrm{min}}\right)^{-4/3} \right)^{1/2} \text{ \quad with } f_\mathrm{min} = 10 \text{ and } f_\mathrm{isco} = \frac{10^3}{\mathcal{M}_\mathrm{det}} \label{eq:bright sirens snr} \\
    p(\hat{\rho}|\rho)
        & = \frac{\exp(-(\rho-\hat{\rho})^2/2)}{\sqrt{2\pi}} \\
    p(\hat{\mathcal{M}}_\mathrm{det}|\mathcal{M}_\mathrm{det})
        & = \frac{\exp(-(\mathcal{M}_\mathrm{det}-\hat{\mathcal{M}}_\mathrm{det})^2/2)}{\sqrt{2\pi}} \\
    p(\hat{z}|z)
        & = \frac{\exp(-(z-\hat{z})^2/2\sigma_{\hat{z}}^2)}{\sqrt{2\pi\sigma_{\hat{z}}^2}} \text{ \quad with } \sigma_{\hat{z}} = 10^{-2} \\
    P(\mathbb{D}|\hat{\rho})
        & = \Theta(\hat{\rho} \geq \hat{\rho}_\mathrm{thr})
\end{align}
Fig.~\ref{fig:bright sirens} shows an estimate for the distance ($\hat{D}$) that is obtained by substituting $\hat{\rho}$ and $\hat{\mathcal{M}}_\mathrm{det}$ into Eq.~\ref{eq:bright sirens snr}.

We simultaneously infer $\mathcal{M}_\mathrm{min}$, $\mathcal{M}_\mathrm{max}$, and $H_0/c$ while fixing $D_\mathrm{max}$.

This model is a somewhat simplified version of the real GW problem in that we only consider a single mass instead of two masses and neglect parameters like orbital inclination.
For simplicity, we also do not assume that the uncertainties scale with $\hat{\rho}$ and model the follow-up noise to always be small (i.e., a host galaxy is always identifiable and proper motion is negligible), which may not be the case~\citep[see, e.g., discussion in ][]{Essick:2024uhl}.


\section{Gaia model}
\label{sec:gaia model}

\begin{figure}
    \begin{center}
        \includegraphics[width=0.49\textwidth]{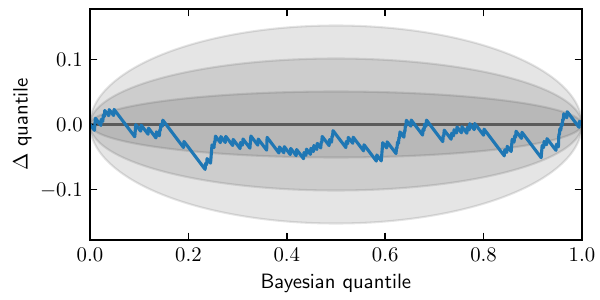}
    \end{center}
    \caption{
        Coverage tests for the Gaia mixture model (Sec.~\ref{sec:rare systems} and Appendix~\ref{sec:gaia model}), analogous to Fig.~\ref{fig:simplified toy model coverage}.
        Because of the higher computational cost associated with this more complex model, we only consider \result{97} catalog realizations and a single follow-up strategy.
        Comparing the measured distribution to a uniform distribution yields a KS $p$-value of \result{0.70}.
    }
    \label{fig:gaia coverage}
\end{figure}

We consider a four dimensional model that consists of a mixture over two subpopulations (contaminants and dark companions).
Each subpopulation consists of the following distributions for the signal to noise ratio ($\rho$), the properties of the system observed by the initial catalog data ($\alpha$ and $\beta$), and the parameter observed by the follow-up that can provide a clear identification of the system type ($\gamma$).
In the case of Gaia binaries, $\rho$ roughly corresponds to the system's magnitude (or brightness), $\alpha$ and $\beta$ could correspond to goodness of fit statistics like \texttt{ruwe}~\citep{Lindegren:2018}, \texttt{amrf}~\citep{Shahaf:2019}, or even just the estimated orbital period, and $\gamma$ corresponds to spectroscopic follow-up that can be decisive in determining the system's composition (i.e., whether there is a bright companion or not).

We denote a (multivariate) normal distribution with mean $\mu$ and covariance matrix $\sigma^2$ as $\mathcal{N}(\mu, \sigma^2)$, and each subpopulation ($a$) is described by
\begin{align}
    p(\rho|\Lambda_a)
        & \propto \rho^{\kappa_\alpha} \Theta(\rho_{\mathrm{min},a} < \rho < \rho_{\mathrm{max},a}) \\
    \alpha, \beta | a
        & \sim \mathcal{N}\left([\mu_{\alpha,a}, \mu_{\beta,a}], \begin{bmatrix} \sigma_{\alpha,a}^2 & \sigma_{\alpha\beta,a}^2 \\ \sigma_{\alpha\beta,a}^2 & \sigma_{\beta,a}^2 \end{bmatrix} \right) \\
    \gamma | a
        & \sim \mathcal{N}(\mu_{\gamma,a}, \sigma_{\gamma,a}^2)
\end{align}
and the total population is
\begin{equation}
    p(\rho, \alpha, \beta, \gamma | \Lambda) = \lambda p(\rho,\alpha,\beta,\gamma|\Lambda_d) + (1 - \lambda) p(\rho, \alpha,\beta,\gamma|\Lambda_c)
\end{equation}
where $\lambda$ is the fraction of systems that contain dark companions ($d$) and $(1-\lambda)$ is the fraction of systems that are contaminants ($c$).
We simultaneously infer 17 hyperparameters: 8 parameters ($\kappa, \mu_\alpha, \mu_\beta, \sigma_\alpha^2, \sigma_\beta^2, \sigma_{\alpha\beta}^2, \mu_\gamma, \sigma_\gamma^2$) for each subpopulation and the mixing fraction ($\lambda$).
We use hyperpriors that require the dark companion population to be the smaller subpopulation ($\lambda < 1/2$).
We also assume that $\alpha$ and $\beta$ are correlated for systems with dark companions but are uncorrelated for contaminants and that $\mu_\gamma$ are typically well-separated between subpopulations so that $\hat{\gamma}$ can fairly conclusively identify the system type if a system is selected for follow-up (i.e., $|\mu_{\gamma,d} - \mu_{\gamma,c}| \gtrsim \sigma_{\gamma,d}, \sigma_{\gamma,c}$).

We additionally model noise by scattering in the maximum likelihood estimates of each property
\begin{align}
    \hat{\rho} 
        & \sim \mathcal{N}(\rho, \sigma_{\hat{\rho}}^2) \\
    \hat{\alpha}, \hat{\beta}
        & \sim \mathcal{N}\left( [\alpha, \beta], \begin{bmatrix} \sigma_{\hat{\alpha}}^2 & \sigma_{\hat{\beta}\hat{y}}^2 \\ \sigma_{\hat{\alpha}\hat{\beta}}^2 & \sigma_{\hat{\beta}}^2 \end{bmatrix} \right) \\
    \hat{\gamma}
        & = \mathcal{N}(\gamma, \sigma_{\hat{\gamma}}^2)
\end{align}
Finally, we model detection as simple thresholds on $\hat{\rho}$, $\hat{\alpha}$, and $\hat{\beta}$.
\begin{equation}
    P(\mathbb{D}|\hat{\rho}, \hat{\alpha}, \hat{\beta}) = \Theta(\hat{\rho} > \hat{\rho}_\mathrm{thr}) \Theta(\hat{\alpha} > \hat{\alpha}_\mathrm{thr}) \Theta(\hat{\beta} > \hat{\beta}_\mathrm{thr})
\end{equation}
which selects for bright events that are estimated to come from a particular part of ($\hat{\alpha}, \hat{\beta}$) parameter space.
Within our experiments, we assumed
\begin{gather}
    \sigma_{\hat{\rho}}^2 = 1 \\
    \sigma_{\hat{\alpha}}^2 = \sigma_{\hat{\beta}}^2 = 0.1 \\
    \sigma_{\hat{\alpha}\hat{\beta}}^2 = 0 \\
    \sigma_{\hat{\gamma}}^2 = 0.01 \\
    \hat{\rho}_\mathrm{thr} = 10 \\
    \hat{\alpha}_\mathrm{thr} = \hat{\beta}_\mathrm{thr} = 0
\end{gather}

For completeness, we also assume $\rho_{\mathrm{min},d}, \rho_{\mathrm{min},c} \ll \hat{\rho}_\mathrm{thr}$ and $\rho_{\mathrm{max},d} = \rho_{\mathrm{max},c} = \infty$ throughout the inference.

Sec.~\ref{sec:gaussian toy model} provides an example derived from this model when we follow-up the \result{100} events that are least likely to be contaminants (largest $\hat{\alpha}^2+\hat{\beta}^2$).
Fig.~\ref{fig:gaia coverage} shows coverage tests (similar to Fig.~\ref{fig:simplified toy model coverage}) with that follow-up strategy.


\bibliography{references}
\bibliographystyle{aasjournalv7}

\end{document}